\documentclass[a4paper,11pt]{article}
\pdfoutput=1

\usepackage{jheppub}
\usepackage{graphicx}
\usepackage{amsmath}
\newcommand{\eq}[1]{eq.~\eqref{eq:#1}}
\newcommand{\eqs}[2]{Eqs.~\eqref{eq:#1} and \eqref{eq:#2}}
\renewcommand{\sec}[1]{section~\ref{sec:#1}}

\newcommand{\subsec}[1]{section~\ref{subsec:#1}}
\newcommand{\subsecs}[2]{sections~\ref{subsec:#1} and \ref{subsec:#2}}
\newcommand{\app}[1]{Appendix~\ref{app:#1}}
\newcommand{\fig}[1]{figure~\ref{fig:#1}}
\newcommand{\figs}[2]{figures~\ref{fig:#1} and \ref{fig:#2}}
\newcommand{\mycites}[1]{refs.~\cite{#1}}
\newcommand{\mycite}[1]{ref.~\cite{#1}}

\newcommand{\Tau}{\mathcal{T}}

\newcommand{\lqcd}{\Lambda_\mathrm{QCD}}

\allowdisplaybreaks[2]

\newcommand{\ord}[1]{{\mathcal O}(#1)}

\newcommand{\ORD}[1]{{\mathcal O}\biggl(#1\biggr)}

\newcommand{\Mae}[3]{\bigl\langle#1\bigr\rvert#2\bigr\rvert#3\bigr\rangle}
\newcommand{\MAe}[3]{\Bigl\langle#1\Bigr\rvert#2\Bigr\rvert#3\Bigr\rangle}

\newcommand{\bnslash}{\bar{n}\!\!\!\slash}
\newcommand{\bare}{\mathrm{bare}}

\newcommand{\bn}{\bar{n}}

\newcommand{\df}{\mathrm{d}}
\newcommand{\img}{\mathrm{i}}

\newcommand{\sdt}{\!\cdot\!}

\newcommand{\Li}{\mathrm{Li}}

\newcommand{\eps}{\epsilon}
\newcommand{\w}{\omega}

\newcommand{\cB}{{\mathcal B}}
\newcommand{\cL}{{\mathcal L}}
\newcommand{\cI}{{\mathcal I}}

\newcommand{\nn}{\nonumber}

\newcommand{\hp}{\hat{p}}
\newcommand{\bnP}{\overline {\mathcal P}}

\newcommand{\conv}{\!\otimes\!}
\newcommand{\convz}{\!\otimes_z\!}

\newcommand{\zero}{{(0)}}
\newcommand{\one}{{(1)}}
\newcommand{\two}{{(2)}}

\newcommand{\tr}{\mathrm{tr}}

\newcommand{\op}{{\mathcal{O}}}

\newcommand{\oq}{{\mathcal{Q}}}

\newcommand{\cusp}{\mathrm{cusp}}
\newcommand{\Disc}{\mathrm{Disc}}

\title{The Quark Beam Function at Two Loops}

\author{Jonathan R.~Gaunt,}
\author{Maximilian Stahlhofen}
\author{and Frank J.~Tackmann}

\affiliation{Theory Group, Deutsches Elektronen-Synchrotron (DESY), D-22607 Hamburg, Germany}

\emailAdd{jonathan.gaunt@desy.de}
\emailAdd{maximilian.stahlhofen@desy.de}
\emailAdd{frank.tackmann@desy.de}

\abstract{
In differential measurements at a hadron collider, collinear initial-state radiation is described by process-independent beam functions. They are the field-theoretic analog of initial-state parton showers. Depending on the measured observable they are differential in the virtuality and/or transverse momentum of the colliding partons in addition to the usual longitudinal momentum fraction. Perturbatively, the beam functions can be calculated by matching them onto standard quark and gluon parton distribution functions. We calculate the inclusive virtuality-dependent quark beam function at NNLO, 
which is relevant for any observables probing the virtuality of the incoming partons, including $N$-jettiness and beam thrust. For such observables, our results are an important ingredient in the resummation of large logarithms at N$^3$LL order, and provide all contributions enhanced by collinear $t$-channel singularities at NNLO for quark-initiated processes in analytic form. We perform the calculation in both Feynman and axial gauge and use two different methods to evaluate the discontinuity of the two-loop Feynman diagrams, providing nontrivial checks of the calculation. As part of our results we reproduce the known two-loop QCD splitting functions and confirm at two loops that the virtuality-dependent beam and final-state jet functions have the same anomalous dimension.
}
\keywords{QCD, NNLO Calculations, Hadronic Colliders}

\begin{document}
%% fix date before submission
{\flushright DESY 14-001\\January 21, 2014\\[-9ex]}
\maketitle

%%%%%%%%%%%%%%%%%%%%%%%%%%%%%%%%%%%%%%%%%%%%%%%%%%%%%%%%%%%%%%%%%%%%%%%%%%%%%%%%
\section{Introduction}
\label{sec:intro}
%%%%%%%%%%%%%%%%%%%%%%%%%%%%%%%%%%%%%%%%%%%%%%%%%%%%%%%%%%%%%%%%%%%%%%%%%%%%%%%%

\begin{figure}
\centering
\includegraphics[scale=0.7]{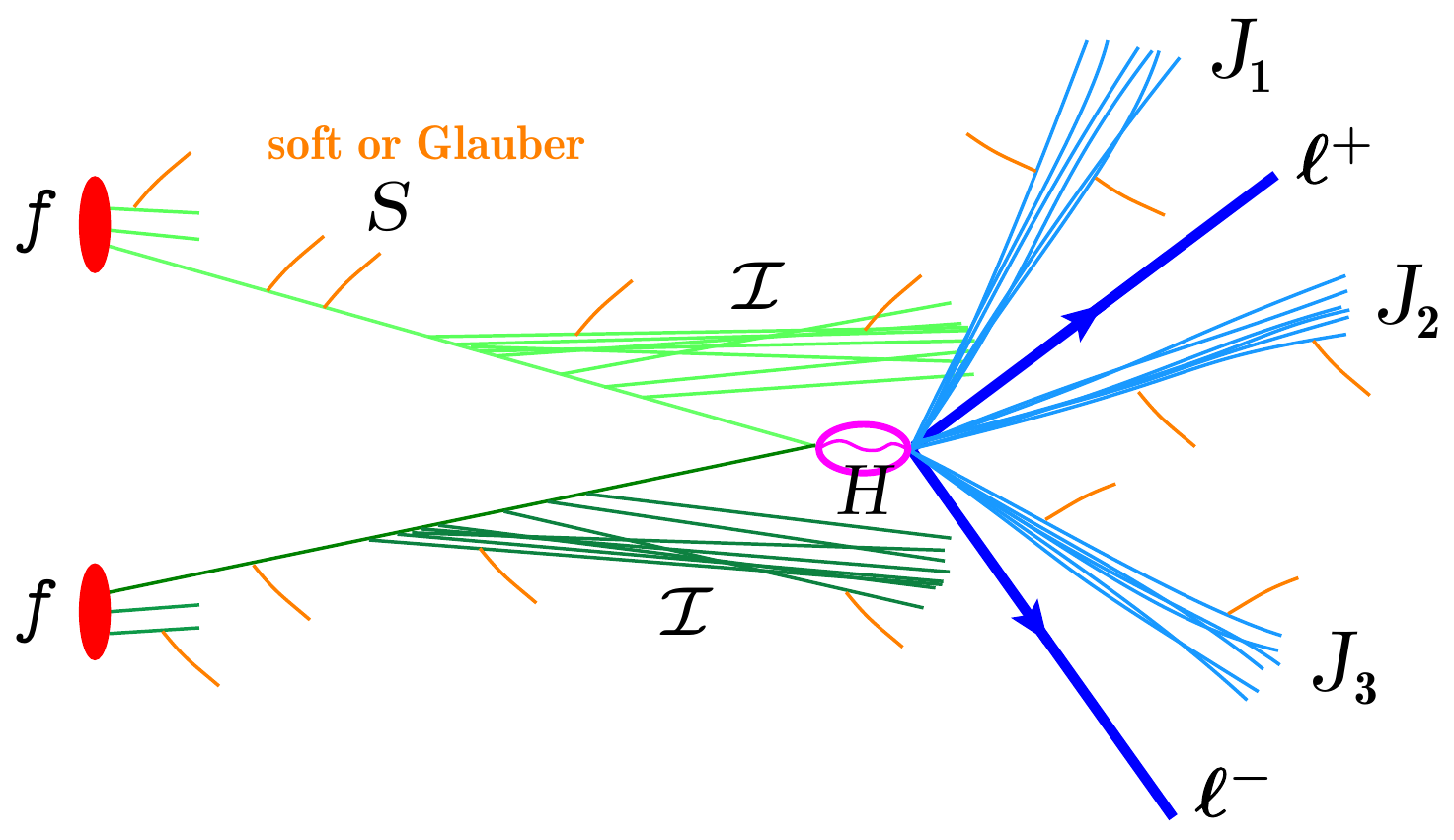}
\caption{Schematic depiction of a generic LHC collision. An energetic parton from each proton emits
initial-state radiation ($\mathcal{I}$) before the two collide in a hard process (${H}$), producing
here a lepton pair and colored particles that emit final state radiation in jets ($J_i$). In addition,
there are soft interactions connecting the initial-state and final-state colored particles ($S$). Also further
pairs of initial-state partons may interact (not shown). Figure taken from \mycite{Stewart:2010qs}.
\label{fig:genLHCcollision}} 
\end{figure}

At the LHC, we are mainly interested in exploring the high-energy, short-distance processes that
produce Higgs particles and possible beyond Standard Model particles. The complication we
face in studying these processes is that they occur inside a QCD environment of initial-state radiation (ISR),
soft interactions, final-state radiation (FSR), and multiparton interactions
due to the the fact that we are colliding bags of colored particles, commonly known as protons.
This is illustrated in \fig{genLHCcollision}.

If we are interested in sufficiently inclusive measurements, for example the total cross section for 
$pp \to L + X$, where $L$ is the colorless final state of interest (e.g. $W$, $Z$, or $H$) and we make no restrictions on
the remaining hadronic final state $X$, then a factorization theorem exists to express the cross section in terms of the
partonic short-distance cross section and the parton distribution functions (PDFs) up to corrections that are suppressed by $\lqcd^2/Q^2$, with $Q^2$ the scale of the hard interaction. In this case, the PDFs describe all of the initial-state radiation effects, and have only two arguments -- the fraction $x$ of light-cone momentum of the proton carried by the parton entering the hard process, and the factorization scale $\mu$~\cite{Bodwin:1984hc, Collins:1985ue, Collins:1988ig}:
%%%
\begin{equation} \label{eq:Xsecbasic}
\df\sigma = \sum_{i,j} \df\sigma_{ij}^{\rm part}(Q,\mu) \conv f_i(x_a,\mu) \conv f_j(x_b,\mu) \biggl[ 1 + \ORD{\frac{\lqcd^2}{Q^2}} \biggr]
\,,\end{equation}
%%%
where $i,j =\{g,u,\bar{u},d,...\}$. The PDFs absorb nonperturbative as well as perturbative effects
and so need to be fitted to data, while their evolution with respect to the scale $\mu$ can be predicted according to the
DGLAP equation.

If one wishes to make more differential measurements, and if the scale of the extra measurement is
substantially lower than the scale of the hard process, then \eq{Xsecbasic} no longer applies
and we either have to derive a new factorization formula, or it is possible that no factorization 
formula exists. If a new factorization formula can be derived, it will have a more complex structure
than \eq{Xsecbasic}, involving separate functions to describe the effects of the soft radiation,
energetic FSR and ISR. In particular, the most appropriate description of energetic ISR may involve
parton density objects that depend on more variables than the standard PDFs, referred
to as beam functions~\cite{Stewart:2009yx} (they are also referred
to as unintegrated PDFs in the literature). The factorization formula differential
in some infrared (IR) sensitive observable $\Tau$ has the generic form
%%%
\begin{equation} \label{eq:genefact}
\dfrac{\df \sigma}{\df \Tau} = H \times \bigl[ B_a \otimes B_b \otimes J_1 \otimes ... \otimes J_n \otimes S \bigr](\Tau)
\end{equation}
%%%
where $\otimes$ denotes a convolution of some sort, and there is a correspondence between the 
objects on the right hand side of \eq{genefact} and the different parts of \fig{genLHCcollision}: $H$ contains
the hard process, while the soft function $S$, the jet functions $J_i$, and the beam functions $B_{a,b}$
describe the contributions to the measurement of $\Tau$ from soft radiation, energetic FSR, and energetic ISR respectively.

A variety of beam functions have appeared in the literature and have been applied in factorization
formulae of various kinds. An extensively studied case~\cite{Collins:1984kg, Catani:2000vq, Collins:1350496, Becher:2010tm, Catani:2010pd, GarciaEchevarria:2011rb, Chiu:2012ir, Echevarria:2012js, Becher:2012yn} is that of the inclusive $p_T$-dependent beam
function $B_i(p_T^2,x,\mu)$, or transverse momentum dependent PDF (TMD PDF), which measures the total
transverse momentum $p_T$ of the parton $i$ entering the hard process in addition to its light-cone momentum fraction $x$. It appears for example in the cross section for $pp \to L + X$ with $L$ uncharged under color in which the transverse momentum of $L$ is measured. We are interested in the case of the inclusive virtuality ($t$)-dependent beam function
$B_i(t, x, \mu)$~\cite{Stewart:2009yx, Stewart:2010qs}. Here, the argument $t$ measures the component of momentum of the parton $i$ in the light-cone direction opposite to the respective proton momentum, or equivalently the (space-like, transverse) virtuality of the parton $i$
entering the hard interaction. The virtuality-dependent beam functions appear in any cross section in which the incoming parton virtualities are probed by the measurement performed on the final state. They have appeared in a variety of processes -- e.g. in $J/\psi$ photoproduction~\cite{Fleming:2006cd}, hadron-hadron $N$-jettiness~\cite{Stewart:2010tn, Jouttenus:2011wh},
including the Higgs and Drell-Yan $0$-jettiness (or beam thrust)~\cite{Stewart:2009yx, Stewart:2010pd, Berger:2010xi} and the Higgs + 1-jet cross section~\cite{Jouttenus:2013hs}, the $1$-jettiness DIS cross section~\cite{Kang:2013nha, Kang:2013lga}, and $1$-jettiness in nuclear dynamics \cite{Kang:2012zr, Kang:2013wca}. There are also more complex beam functions, that depend on both $p_T$ and $t$ \cite{Collins:2007ph, Rogers:2008jk, Mantry:2009qz, Mantry:2010mk, Mantry:2010bi, Jain:2011iu, Kang:2013nha}, or that involve more complicated jet-algorithm dependent measurements~\cite{Becher:2012qa, Tackmann:2012bt, Liu:2012sz, Becher:2013xia, Stewart:2013faa, Li:2014ria}.

The standard PDFs can be defined as proton matrix elements of renormalized operators $\oq_i$ composed of partonic fields,
%%%
\begin{align} \label{eq:f_def}
f_i(x, \mu) &= \Mae{p_n(P^-)}{\oq_i(x P^-, \mu)}{p_n(P^-)}
\,,\end{align}
%%%
where $\lvert p_n(P^-)\rangle$ denotes the external proton state with lightlike momentum $P^\mu=P^- n^\mu/2$. The matrix elements are always averaged over proton spins, which we suppress in our notation. Similarly, a generic beam function $B(x,\Tau,\mu)$ that depends on the extra dimensionful quantity $\Tau$ (where $\Tau = p_T^2, t, \ldots$) can be expressed in soft-collinear effective theory (SCET)~\cite{Bauer:2000ew,Bauer:2000yr,Bauer:2001ct,Bauer:2001yt,Bauer:2002nz,Beneke:2002ph} as the proton matrix elements of renormalized operators composed of partonic fields with an additional dependence on $\Tau$,
%%%
\begin{align} \label{eq:B_def}
B_i(\Tau, x, \mu) &= \Mae{p_n(P^-)}{\op_i(\Tau, xP^-, \mu)}{p_n(P^-)}
\,.\end{align}
%%%
Like the PDFs, the $B_i$ contain nonperturbative information and cannot be predicted
entirely from theory. We could fit them from data just like the PDFs -- however, this would be a very difficult 
undertaking due to the paucity of appropriate data and the fact that they depend on a further argument $\Tau$.
However, as long as we are in a region in which $\Tau$ is large enough, i.e. $\Tau \gg \lqcd^2$, we can exploit the fact that in this region $\Tau$ is predominantly generated perturbatively (up to corrections suppressed by powers of $1/\Tau$). In this region, the beam functions can be calculated as the convolution of the usual collinear PDFs, which provide the nonperturbative information, and a perturbatively calculable matching coefficient $\mathcal{I}_{ij}(\Tau, x, \mu)$ that describes the $\Tau$ dependence due to perturbative ISR:
%%%
\begin{equation} \label{eq:genematch}
B_i(\Tau,x,\mu) = \sum_j \mathcal{I}_{ij}(\Tau,x,\mu) \otimes f_j(x,\mu) \biggl[ 1 + \ORD{\frac{\lqcd^2}{\Tau}} \biggr]
\,.\end{equation}
%%%
This equation can be regarded as an operator product expansion (OPE) in SCET~\cite{Fleming:2006cd, Stewart:2009yx}.

In almost all of the above-mentioned examples, the matching coefficients are known at next-to-leading order (NLO) in perturbation theory. For the inclusive $p_T$-dependent beam function or TMD PDF, the coefficients at next-to-next-to-leading order (NNLO) have been obtained in \mycite{Catani:2013tia} using the NNLO calculations in \mycites{Catani:2011kr, Catani:2012qa} (except for gluon-spin-correlated contributions). The quark-to-quark $\cI_{qq}(p_T^2, x, \mu)$ matching coefficient has been directly calculated at NNLO using SCET in \mycite{Gehrmann:2012ze}, and the remaining partonic channels are being calculated~\cite{Gehrmann:2014uaa}.

We are interested in the inclusive virtuality-dependent beam function, which for brevity in the following we refer to simply as the beam function. It has been calculated at NLO in \mycites{Stewart:2010qs, Berger:2010xi}. In this paper, we perform the two-loop calculation for the quark beam function in SCET, i.e. we calculate the matching coefficients for the quark and antiquark beam functions, $\cI_{qj}(t, x, \mu)$ and $\cI_{\bar q j}(t, x, \mu)$, at NNLO. They are relevant for processes where the hard interaction is initiated by quarks. The gluon beam function at NNLO will be presented in \mycite{Gaunt:2014cfa}.
The NNLO beam functions are a necessary ingredient for the N$^3$LL prediction of the $N$-jettiness cross section and all other observables involving the virtuality-dependent beam functions. From a fixed-order perspective, the NNLO beam function describes the full two-loop singular contributions differential in $t$ and $x$ from collinear initial-state $t$-channel singularities.

This paper is organized as follows. In \sec{match} we discuss the general setup of the matching calculation, including the relevant operator definitions and renormalization group equations (RGEs). For the calculation itself, we utilize two different methods, which are described in \sec{method}. Our results for the NNLO quark and antiquark matching coefficients $\mathcal{I}_{qj}$ and $\mathcal{I}_{\bar qj}$ are presented in \sec{results}. 
We conclude in \sec{conclusions}.

%%%%%%%%%%%%%%%%%%%%%%%%%%%%%%%%%%%%%%%%%%%%%%%%%%%%%%%%%%%%%%%%%%%%%%%%%%%%%%%%
\section{General setup and matching at NNLO}
\label{sec:match}
%%%%%%%%%%%%%%%%%%%%%%%%%%%%%%%%%%%%%%%%%%%%%%%%%%%%%%%%%%%%%%%%%%%%%%%%%%%%%%%%

Our calculation of the (virtuality-dependent) quark beam function follows the general setup in \mycite{Stewart:2010qs}. For a detailed discussion we refer the reader there. Here, we summarize the relevant definitions of the bare and renormalized beam function and PDF operators in \subsec{Bdef}, and in \subsec{matchdet} we explain the necessary steps to extract the matching coefficients $\mathcal{I}_{ij}$ and give all relevant expressions at NNLO.

%===============================================================================
\subsection{Operator definitions of beam functions and PDFs}
\label{subsec:Bdef}
%===============================================================================

We use the usual light-cone (Sudakov) decomposition for four-vectors, writing an arbitrary four-vector $A^\mu$ as
%%%
\begin{equation} \label{eq:LCSCET}
A^\mu = A^- \frac{n^\mu}{2} + A^+ \frac{\bn^\mu}{2} + A_\perp^\mu \, ,
\end{equation}
%%%
with
%%%
\begin{align}
 n^2 = \bn ^2 = 0\,, \quad n \cdot \bn = 2
 \,,\qquad
A^+ = n\cdot A
\,, \quad
A^- = \bn\cdot A
\,, \quad
n\cdot A_\perp = \bn\cdot A_\perp = 0
\,.\end{align}
%%%
Quarks and gluons with momentum $p$ are $n$-collinear if their momentum components scale  as $(p^+, p^-, p_\perp) \sim p^-(\lambda^2, 1, \lambda)$, where $\lambda \ll 1$ is the power expansion parameter of SCET. The effective size of $\lambda$ is set by the measurement of interest, e.g., $\lambda^2 \simeq \mathcal{T}_N/Q$ for $N$-jettiness~\cite{Stewart:2010tn}.

In SCET, $n$-collinear quarks and gluons are described by composite quark and gluon field operators $\chi_n$ and $\cB_{n\perp}^\mu$,
%%%
\begin{align} \label{eq:chiSCET}
\chi_n(y) = W_n^\dagger(y)\, \xi_n(y)
\,, \qquad
\cB_{n\perp}^\mu(y) = \frac{1}{g} \bigl[W_n^\dagger(y)\, \img D_{n\perp}^\mu W_n(y) \bigr]
\,,\end{align}
%%%
where $\xi_n$ is the $n$-collinear quark field and $\img D_{n\perp}^\mu = \mathcal{P}_{n\perp}^\mu + g A_{n\perp}^\mu$ is the covariant derivative involving the $n$-collinear gluon field $A_{n}^\mu$. The Wilson line
%%%
\begin{align} \label{eq:Wn}
W_n(y) = \biggl[\sum_\text{perms} \exp\Bigl(-\frac{g}{\bnP_n}\,\bn\sdt A_n(y)\Bigr)\biggr]
\end{align}
%%%
accounts for arbitrary emissions of $n$-collinear gluons, which are all of $\ord{\lambda^0}$ in the power counting. The $\bnP_n$ and $\mathcal{P}_{n\perp}$ are the SCET label momentum operators~\cite{Bauer:2001ct}.

The $\chi_n$ and $\cB_{n\perp}^\mu$ fields are gauge invariant with respect to collinear gauge transformations~\cite{Bauer:2000yr, Bauer:2001ct}. They are defined after the field redefinition~\cite{Bauer:2001yt} decoupling soft gluons from collinear particles and do not interact with soft gluons through their Lagrangian (at leading order in the power counting). This means they do not transform under soft gauge transformation, such that operators built from them are gauge invariant under both soft and collinear gauge transformations. The soft interactions with collinear particles are factorized into a product of soft Wilson lines, whose matrix element gives the soft function in \eq{genefact}.

The bare quark, antiquark, and gluon beam function operators are defined in terms of the fields in \eq{chiSCET} as~\cite{Stewart:2010qs}%
\footnote{Note that the beam function definition in \mycite{Stewart:2010qs} contains an explicit $\theta(\w)$, which we have chosen to move into the definition of the operators $\op_i$, as was done in \mycite{Berger:2010xi}.}
\begin{align} \label{eq:BOpdefSCET}
\op^\bare_q(t, \omega)
&= \theta(\w)\, {\bar \chi_n(0)\, \delta(t - \w\hp^+)\, \frac{\bnslash}{2} \bigl[\delta(\w - \bnP_n) \chi_n(0)\bigr]}
\,, \nn \\
\op_{\bar q}^\bare(t,\w)
&= \theta(\w)
\tr \Bigl\{\frac{\bnslash}{2} \chi_n (0) \, \delta(t - \w\hp^+) \,\bigl[\delta(\w - \bnP_n) \bar\chi_n(0)\bigr]\Bigr\}
\,, \nn \\
\op^\bare_g(t, \omega) &= -\w\,\theta(\w) \, \cB_{n\perp\mu}^c(0)\, \delta(t - \w\hp^+) \bigl[\delta(\w - \bnP_n) \cB_{n\perp}^{\mu c}(0) \bigr]
\,.\end{align}
%%%
Here, $\bnP_n$ acts within the square brackets and returns the sum of the large minus momentum components of all fields in $\chi_n$ and $\cB_{n\perp}$. Hence, the effect of the $\delta(\w - \bnP_n)$ operator is to set the total minus momentum of the composite quark/gluon field to $\omega$. The $\hp^+$ is the momentum operator of the small plus momentum, and the effect of $\delta(t - \w\hp^+)$ is to set the total plus momentum of all initial-state radiation (i.e. of any intermediate state inserted between the fields) to $t/\w$.

The bare quark, antiquark, and gluon PDF operators are defined in terms of collinear SCET fields as
%%%
\begin{align} \label{eq:oq_def}
\mathcal{Q}^\bare_q(\w)
&= \theta(\w)\, \bar{\chi}_n(0) \frac{\bnslash}{2} \bigl[\delta(\w - \bnP_n) \chi_n(0)\bigr]
\,, \nn \\
\mathcal{Q}^\bare_{\bar q}(\w)
&= \theta(\w)\, \tr \Bigl\{\frac{\bnslash}{2} \chi_n(0) \bigl[\delta(\w - \bnP_n) \bar\chi_n(0)\bigr] \Bigr\}
\,, \nn \\
\mathcal{Q}^\bare_g(\w)
& = -\w\theta(\w)\, \cB_{n\perp\mu}^c(0) \bigl[\delta(\w - \bnP_n) \cB_{n\perp}^{\mu c}(0) \bigr]
\,.\end{align}
%%%
In contrast to the operators in \eq{BOpdefSCET}, they only measure the large minus components of the fields.

The beam functions and PDFs are defined as in \eqs{B_def}{f_def} as the proton matrix elements of the corresponding renormalized operators $\op_i$ and $\mathcal{Q}_i$, which are related to the bare operators by
%%%
\begin{align} \label{eq:Oop_ren}
\mathcal{O}_i^\bare(t, \w)
&= \int\! \df t'\, Z_B^i(t - t', \mu)\, \mathcal{O}_i(t', \w, \mu)
\,,\\ \label{eq:Qop_ren}
\mathcal{Q}_i^\bare(\w)
&= \sum_j \int\! \frac{\df \w'}{\w'}\, Z^f_{ij}\Big(\frac{\w}{\w'},\mu\Big) \mathcal{Q}_j(\w',\mu)\,.
\end{align}
%%%
Here and in the following we never sum over repeated parton indices unless explicitly stated otherwise.
The renormalization constants $Z_B^i(t - t', \mu)$ and $Z^f_{ij}(\w/\w',\mu )$ are defined to remove the ultraviolet (UV) divergences in the bare beam function and PDF matrix elements, respectively. 
In this paper, we always renormalize using the $\overline{\mathrm{MS}}$ scheme with dimensional regularization (which yields
the standard renormalized PDFs). The renormalization of the PDF operators is well known. In contrast to the PDFs,
the renormalization of the beam function operator in SCET depends on $t$ but not on $\w$. This structure
was proven to all orders in perturbation theory in \mycite{Stewart:2010qs}, with the renormalization constant
(or equivalently the anomalous dimension) being identical to that of the SCET jet function.

As already mentioned, the collinear fields in the operators include a field redefinition decoupling them from soft interactions. Hence, their matrix elements (at leading order in the power counting) are computed using the purely $n$-collinear sector of the SCET Lagrangian, which has the same form as the full QCD Lagrangian in a boosted frame. Therefore, as is well known, the matrix elements of boost-invariant collinear operators, such as those of the beam functions and PDFs, can be computed using QCD Feynman rules. This renders the calculation more compact because it avoids having to use the more complex SCET vertices.

As discussed in detail in \mycite{Stewart:2010qs}, the definitions of the PDFs in terms of SCET fields are equivalent to the standard operator definitions in QCD. For example, the quark PDF in terms of the full QCD quark field $\psi(x)$ in position space is defined as~\cite{Collins:1981uw}
%%%
\begin{equation} \label{eq:f_def_QCD}
f_q(\w/P^-, \mu) = \theta(\w) \int\! \frac{\df y^+}{4\pi}\,
  e^{-\img \w y^+/2}
  \MAe{p_n(P^-)}{\Bigl[\bar\varphi\Bigl(y^+\frac{\bn}{2}\Bigr)
   \frac{\bnslash}{2} \varphi(0) \Bigr]_\mu}{p_n(P^-)}
\,,\end{equation}
%%%
where we used square brackets to denote the renormalized operator and $\varphi(x)$ comprises a quark field attached to a lightlike Wilson line:
%%%%
\begin{equation} \label{eq:GIquark}
 \varphi(y)
= P\biggl\{\exp \biggl[-\img g \int_0^\infty \frac{\df x^+}{2}\,
\bn\cdot A\Bigl(y + x^+ \frac{\bn}{2}\Bigr) \biggr] \biggr\}\psi(y)
\,.\end{equation}
%%%
The inclusion of the Wilson line renders the product of fields separated
along the $\bn$ direction in \eq{f_def_QCD} gauge invariant. The SCET fields in \eq{oq_def}
involve a Fourier transform of $\psi$ in $y^+$, and the Wilson line in \eq{GIquark}
corresponds to the $W_n$ contained in the definitions of $\chi_n$
and $\cB_{n\perp}^\mu$. In general, the definition of the $n$-collinear SCET fields explicitly excludes the soft region where all their momentum components become small.
In practice, this is implemented by carrying out all integrations in the calculation of collinear SCET matrix elements over the full momentum range (including the soft region) and then subtracting the zero-bin contributions where the collinear momenta become soft~\cite{Manohar:2006nz}. As is well known, the soft region does not contribute to the PDFs, which means that no zero-bin subtractions are needed for the collinear SCET fields appearing in the PDF operators. Therefore, as long as one uses the same renormalization scheme, the QCD and SCET definitions of the PDF are precisely equivalent at both the calculational
and formal level.

The situation for the beam function operators is more complicated. First, the explicit dependence on the small $\hp^+$ momentum corresponds to a large separation in $y^-$ in position space (as can be seen from the corresponding definitions in \mycite{Stewart:2010qs}). Thus the quark beam function operator expressed in terms of full QCD fields should contain a quark and an antiquark field operator separated in both the $n$ and $\bn$ directions, Fourier-transformed with respect to both of these separations. Since the fields have to be separated in both light-cone directions, it is a priori not clear how to obtain an unambiguous gauge-invariant definition for the beam function in terms of full QCD fields, because different paths for the Wilson lines connecting the fields are not equivalent. Second, the zero-bin subtractions on the collinear SCET fields are required in the case of the beam function operators. Practically, they involve scaleless integrals and vanish. Formally, the scaleless zero-bin subtraction is still crucial as it converts $1/\eps$ soft IR divergences into $1/\eps$ UV divergences that can be renormalized.
Nevertheless it is interesting to note that performing the calculation of the beam function matrix elements in SCET using QCD Feynman rules (with a vanishing zero-bin subtraction) is operationally equivalent to evaluating the matrix elements of the QCD operator
%%%
\begin{align} \label{eq:B_def_QCD}
\theta(\w)\, \dfrac{1}{\w} \int\! \frac{\df y^+}{4\pi} \frac{\df y^-}{4\pi}
  e^{-\img \frac{\w y^+}{2} + \img \frac{y^-t}{2\w} }{\Bigl[\bar\varphi \Bigl(y^+\frac{\bn}{2}+y^-\frac{n}{2}\Bigr)
    \frac{\bnslash}{2} \varphi(0) \Bigr]}
\,.\end{align}
%%%

%===============================================================================
\subsection{Calculation of the quark matching coefficients}
\label{subsec:matchdet}
%===============================================================================

The explicit form of the matching equation for the beam functions is~\cite{Stewart:2010qs}
%%%
\begin{align} \label{eq:BOPE}
B_{i}(t,x,\mu) &= \sum_j  \int \! \dfrac{\df z}{z}\, \cI_{ij}(t,z,\mu)\, f_{j}\Bigl(\frac{x}{z},\mu \Bigr)
\bigg[1 + \ORD{\frac{\lqcd^2}{t}} \bigg]
\,.\end{align}
%%%
Our aim is to calculate the quark matching coefficients $\cI_{qj}$ in \eq{BOPE} at $\mathcal{O}(\alpha_s^2)$.
Since the $\cI_{ij}$ only depend on short-distance physics they can be calculated in perturbation theory at the scale $\mu^2\sim t$. Furthermore, the matching holds at the operator level in the sense that it is independent of the external state, so to perform the matching calculation we can choose whatever external state is most convenient, provided that state has some overlap with the parton state $j$~\cite{Bardeen:1978yd}. We choose to use on-shell massless quarks and gluons that are $n$-collinear with momentum $p^\mu = p^- n^\mu/2$ as the external states. The corresponding partonic beam functions and PDFs with parton $j$ in the external state are denoted as
%%%
\begin{align} %\label{eq:partonicB_def}
B_{i/j}(t,z,\mu) &= \Mae{j_n(p^-)}{\op_i(t,\w,\mu)}{j_n(p^-)}
\,, \nn \\
%\label{eq:partonicf_def}
f_{i/j}(z,\mu) &= \Mae{j_n(p^-)}{\oq_i(\w,\mu)}{j_n(p^-)}
\,,\end{align}
%%%
and analogously for the bare versions. Here, $z=\w/p^-$ is now defined as the fraction of parton $j$'s minus momentum carried by parton $i$. The matrix elements are always understood to be averaged over the color and spin of parton $j$. 

In terms of the partonic matrix elements, the matching equation becomes
%%%
\begin{align} \label{eq:Bmatching}
B_{i/j}(t,z,\mu) &= \sum_k  \int \! \dfrac{\df z'}{z'}\, \cI_{ik}(t,z',\mu)\, f_{k/j}\Bigl(\frac{z}{z'},\mu \Bigr)
\nn \\
& \equiv \sum_k \cI_{ik}(t,z,\mu) \convz f_{k/j}(z,\mu)
\,,\end{align}
%%%
where in the second line we defined the shorthand notation $\convz$ to denote the Mellin convolution in the light-cone minus component and the integration limits are implicit in the support of the functions, $z' < 1$ and $z/z' < 1$.

The partonic matrix elements $B_{i/j}$ and $f_{i/j}$ in \eq{Bmatching} are renormalized quantities. However, unlike the proton matrix elements $B_i$ and $f_i$, they are unphysical and only serve as a tool to perform the matching. Using on-shell external states makes the calculation simpler algorithmically because there are fewer scales involved. This comes at the cost of introducing explicit IR divergences in both $B_{i/j}$ and $f_{i/j}$ due to which they are ill-defined in four dimensions even after UV renormalization without further IR regularization. We use dimensional regularization in $d = 4 - 2\epsilon$ dimensions to also regulate the IR divergences alongside the UV divergences. Therefore, we need to perform the entire calculation in $d = 4 - 2\epsilon$  dimensions, right up until the point at which we obtain the matching coefficient $\cI_{ik}$ [see \eq{Bmatching2} below]. We then set $d = 4$ to obtain the final result for $\cI_{ik}$.

Our two-loop calculation (see \sec{method}) yields the $\mathcal{O}(\alpha_s^2)$ piece of the bare partonic quark beam function $B^{\bare}_{q/j}$. The bare and renormalized partonic matrix elements are related through the renormalization constants $Z_B^j$ in \eq{Oop_ren},
%%%
\begin{align} \label{eq:Brenorm}
B^{\mathrm{bare}}_{i/j} (t,z,g_0) = \int \df t'\, Z_B^i(t-t',\mu,g)  B_{i/j}(t',z,\mu,g)
\,.\end{align}
%%%
Order by order in the perturbative expansion, $Z_B^i(t, \mu, g)$ can be derived from the known beam function anomalous dimension $\gamma_B^i(t,\mu)$~\cite{Stewart:2010qs}. Here, we have explicitly denoted the dependence on the bare and renormalized coupling, $g_0$ and $g$, for which we need the one-loop relation
%%%
\begin{align}
g_0 = g\biggl(\dfrac{\mu^2 e^{\gamma_E}}{4\pi}\biggr)^{\epsilon/2}
\Bigl[1-\dfrac{\alpha_s}{8\pi\epsilon} \beta_0 +\mathcal{O}(\alpha_s^2) \Bigr]
\,,\end{align}
%%%
with $\beta_0 = (11 C_A - 4 T_F n_f)/3$ for $n_f$ active flavors.

We are interested in the part of \eq{Brenorm} proportional to $\alpha_s^2$. Let us
define the perturbative expansions of $B_{i/j}$ (either bare or renormalized) and $Z_B^{i}$ as
%%%
\begin{align}
B_{i/j} = \sum_{n} \left(\dfrac{\alpha_s}{4\pi}\right)^n B_{ij}^{(n)}
\,, \qquad
Z^{i}_{B} = \sum_{n} \left(\dfrac{\alpha_s}{4\pi}\right)^n Z_B^{i(n)}
\,.\end{align}
%%%
Then the term in $B_{i/j}$ we are going to compute is $B^{(2)}_{i/j}$, and \eq{Brenorm} gives
%%%
\begin{align} \label{eq:Brenorm2}
B^{\mathrm{bare},(2)}_{i/j}(t,z) = Z_B^{i(2)}(t,\mu) \, \delta_{ij} \, \delta(1\!-\!z) + B^{(2)}_{i/j}(t,z,\mu)
%\\ \nn
+ \int \!\df t' Z_B^{i(1)}(t\!-\!t',\mu) B^{(1)}_{i/j}(t',z,\mu)
\,.\end{align}
%%%

Having obtained $B^{(2)}_{q/j}$ from \eq{Brenorm2}, we can extract the two loop $\cI^{(2)}_{qj}$ by expanding
\eq{Bmatching} to second order in $\alpha_s$. Let us define the expansions of $f_{i/j}$ and $\cI_{ij}$ as
%%%
\begin{align}
f_{i/j} = \sum_{n=0}^{\infty} \left(\dfrac{\alpha_s}{2\pi}\right)^n f_{i/j}^{(n)}
\,, \qquad
\cI_{ij} = \sum_{n=0}^{\infty} \left(\dfrac{\alpha_s}{4\pi}\right)^n \cI_{ij}^{(n)}
\,.\end{align}
%%%
Then, the part of \eq{Bmatching} proportional to $\alpha_s^2$ yields the matching relation
%%%
\begin{align} \label{eq:Bmatching2}
\cI_{ij}^{(2)}(z,t,\mu) =
B_{i/j}^{(2)}(z,t,\mu) - 4f_{i/j}^{(2)}(z,\mu)\delta(t) - 2\sum_k\cI^{(1)}_{ik}(z,t,\mu) \convz f^{(1)}_{k/j}(z,\mu)
\,.\end{align}
%%%
All individual terms on the right-hand side are IR divergent, and as mentioned earlier, have to be consistently evaluated in $d$ dimensions. In particular, in the last term we need the one-loop contribution to $\cI_{ij}$ in $d = 4 - 2\epsilon$ dimensions, which can be straightforwardly obtained from Appendix C of \mycite{Stewart:2010qs}. The IR divergences cancel between the terms such that we obtain an IR-finite result for $\cI_{ij}^{(2)}(z,t,\mu)$, and we can take the limit $\eps\to 0$.

The renormalized PDF matrix elements $f_{i/j}(z, \mu)$ are related to the bare ones by [see \eq{Qop_ren}]
%%%
\begin{align}
f^{\rm bare}_{i/j}(z) = \sum_k \int\! Z^f_{ik}(z, \mu) \convz f_{k/j}(z, \mu)
\,.\end{align}
%%%
In pure dimensional regularization, all loop corrections to the bare partonic PDF matrix elements are scaleless and vanish.
Hence, the $\overline{\text{MS}}$ renormalized $f^{(n)}_{i/j}$ (for $n \ge 1$) are given by a pure counterterm contribution. In particular
$f^{(1)}_{i/j}$ and $f^{(2)}_{i/j}$, which are needed in \eq{Bmatching2}, are expressed in terms of the well-known one- and two-loop splitting functions $P^{(0)}_{ij}$ and $P^{(1)}_{ij}$ as
%%%
\begin{align} \label{eq:PDFoneloop}
f^{(1)}_{i/j}(z)
&= - \frac{1}{\epsilon}P^\zero_{ij}(z)
\,, \\ \label{eq:PDFtwoloop}
f^{(2)}_{i/j}(z)
&= \frac{1}{2\epsilon^2} \sum_k P^\zero_{ik}(z) \convz P^\zero_{kj}(z) + \frac{\beta_0}{4\epsilon^2}P^\zero_{ij}(z)
- \frac{1}{2\epsilon}P^\one_{ij}(z)
\,.\end{align}
%%%

Note that since $f^{(2)}_{i/j}$ only contains $1/\eps^n$-poles and $\cI^{(2)}_{ij}$ contains none, knowing $B^{(2)}_{q/j}$
and the one-loop quantities in \eq{Bmatching2} allows us to extract both $\cI^{(2)}_{qj}$ and
$f^{(2)}_{q/j}$, the latter of which allows us to calculate the two-loop splitting functions $P^{(1)}_{qj}$ via \eq{PDFtwoloop}.
Hence, from our calculation we get an independent determination of $P^{(1)}_{qj}$ ``for free'', which should of course agree with the known results~\cite{Curci:1980uw, Furmanski:1980cm, Ellis:1996nn}, serving as a very useful cross check of our calculation. Formally, the fact that the beam function calculation reproduces the complete set of IR divergences in \eqs{PDFoneloop}{PDFtwoloop}, such that it yields an IR-finite result for $\cI_{ij}$, shows that SCET correctly reproduces the IR structure of QCD at two loops as it should. Since we use dimensional regularization for both UV and IR, this statement relies on knowing the UV renormalization of the beam function. Alternatively, we can take it for granted that SCET reproduces the IR structure of QCD, in which case our calculation provides an explicit confirmation at two loops that the beam function renormalization is identical to that of the jet function. (Having both checks at the same time would require using different regulators in the UV and IR in order to separate UV and IR divergences as was done at one loop in \mycite{Stewart:2010qs}.)

The $\mu$-dependent logarithmic terms in $\cI^{(2)}_{ij}(t, z, \mu)$ can be obtained from the known two-loop RGE, as has been done for the $j = g$ case in \mycite{Berger:2010xi}. The procedure is as follows. We start from the all-order RGEs for the beam functions,
%%%
\begin{align} \label{eq:B_RGE}
\mu \frac{\df}{\df \mu} B_i(t, x, \mu) &= \int\! \df t'\, \gamma_B^i(t-t',\mu)\, B_i(t', x, \mu)
\,, \nn \\
\gamma_B^i(t, \mu)
&= -2 \Gamma^i_{\cusp}[\alpha_s(\mu)]\,\frac{1}{\mu^2}\cL_0\Bigl(\frac{t}{\mu^2}\Bigr) + \gamma_B^i[\alpha_s(\mu)]\,\delta(t)
\,,\end{align}
%%%
and the PDFs,
%%%
\begin{equation} \label{eq:f_RGE}
\mu\frac{\df}{\df\mu} f_i(z,\mu)
= \sum_j 2 P_{ij} [z,\alpha_s(\mu)] \convz f_j(z,\mu)
\,,\end{equation}
%%%
where
%%%
\begin{align}
\Gamma^i_\cusp(\alpha_s) = \sum_{n=0}^\infty \Gamma^i_n \Bigl(\frac{\alpha_s}{4\pi}\Bigr)^{n+1}
\,, \qquad
\gamma^i_B(\alpha_s) = \sum_{n=0}^\infty \gamma_{B\,n}^i \Bigl(\frac{\alpha_s}{4\pi}\Bigr)^{n+1}
\,,\end{align}
%%%
are the cusp anomalous dimension and the noncusp part of the beam function anomalous dimension, respectively.
The expansion of the splitting functions $P_{ij}$ in terms of the $P_{ij}^{(n)}$ is given in \eq{Pijexp}.
From eqs.~\eqref{eq:B_RGE},~\eqref{eq:f_RGE}, and~\eqref{eq:BOPE} we can derive the RGE for the Wilson coefficients $\cI_{ij}(t, z, \mu)$,
%%%
\begin{align}
\mu\frac{\df}{\df\mu}\cI_{ij}(t,z,\mu)
= \sum_k \int\!\df t'\,  \cI_{ik}\Bigl(t - t', z,\mu\Bigr) \convz \Bigl[\gamma_B^i(t', \mu)\, \delta_{kj} \delta(1-z)
- 2 \delta(t') P_{kj}(z,\mu) \Bigr]
\,.\end{align}
%%%
Solving this equation recursively to two-loop order we obtain
%%%
\begin{align} \label{eq:I2master}
\cI_{ij}^\two(t,z,\mu)
&= \frac{1}{\mu^2} \cL_3\Bigl(\frac{t}{\mu^2}\Bigr) \frac{(\Gamma_0^i)^2}{2}\, \delta_{ij}\delta(1-z)
  \nn \\ & \quad
  + \frac{1}{\mu^2} \cL_2\Bigl(\frac{t}{\mu^2}\Bigr)
  \Gamma_0^i \Bigl[- \Bigl(\frac{3}{4} \gamma_{B\,0}^i + \frac{\beta_0}{2} \Bigr) \delta_{ij}\delta(1-z) + 3P^\zero_{ij}(z) \Bigr]
  \nn \\ & \quad
  + \frac{1}{\mu^2} \cL_1\Bigl(\frac{t}{\mu^2}\Bigr)
  \biggl\{ \Bigl[\Gamma_1^i - (\Gamma_0^i)^2 \frac{\pi^2}{6} + \frac{(\gamma_{B\,0}^i)^2}{4} + \frac{\beta_0}{2} \gamma_{B\,0}^i \Bigr] \delta_{ij}\delta(1-z)
  \nn \\ & \qquad
  + 2\Gamma_0^i\, I^\one_{ij}(z)
  - 2 (\gamma_{B\,0}^i + \beta_0) P^\zero_{ij}(z)
  + 4 \sum_k P^\zero_{ik}(z)\conv_z P^\zero_{kj}(z) \biggr\}
  \nn \\ & \quad
  + \frac{1}{\mu^2} \cL_0\Bigl(\frac{t}{\mu^2}\Bigr)
  \biggl\{ \Bigl[(\Gamma_0^i)^2 \zeta_3 + \Gamma_0^i \gamma_{B\,0}^i \frac{\pi^2}{12} - \frac{\gamma_{B\,1}^i}{2} \Bigr]
  \delta_{ij} \delta(1-z)
  - \Gamma_0^i \frac{\pi^2}{3} P^\zero_{ij}(z)
  \nn \\ & \qquad
  - (\gamma_{B\,0}^i + 2\beta_0) I^\one_{ij}(z)
  + 4 \sum_k I^\one_{ik}(z)\conv_z P^\zero_{kj}(z)
  + 4 P^\one_{ij}(z)
  \biggr\}
  \nn \\ & \quad
  + \delta(t)\, 4 I^\two_{ij}(z)\,,
\end{align}
%%%
where we denote the plus distributions as
%%%
\begin{align} \label{eq:plusdefmaintxt}
\cL_n(x)
&= \biggl[ \frac{\theta(x) \ln^n x}{x}\biggr]_+
 = \lim_{\eps \to 0} \frac{\df}{\df x}\biggl[ \theta(x- \eps)\frac{\ln^{n+1} x}{n+1} \biggr]\,.
\end{align}
%%%
All terms on the right hand side of \eq{I2master} are known apart from the two-loop matching functions $I^\two_{ij}(z)$, which are the novel output of our calculation performed in this paper. We present our results for the quark $I^\two_{qj}(z)$ functions in \sec{results}. All other quantities in \eq{I2master} contributing for $i=q$ are collected in \app{Formulae}.

%%%%%%%%%%%%%%%%%%%%%%%%%%%%%%%%%%%%%%%%%%%%%%%%%%%%%%%%%%%%%%%%%%%%%%%%%%%%%%%%
\section{Method of calculation}
\label{sec:method}
%%%%%%%%%%%%%%%%%%%%%%%%%%%%%%%%%%%%%%%%%%%%%%%%%%%%%%%%%%%%%%%%%%%%%%%%%%%%%%%%

\begin{figure}[t]
\includegraphics[width=0.245\textwidth]{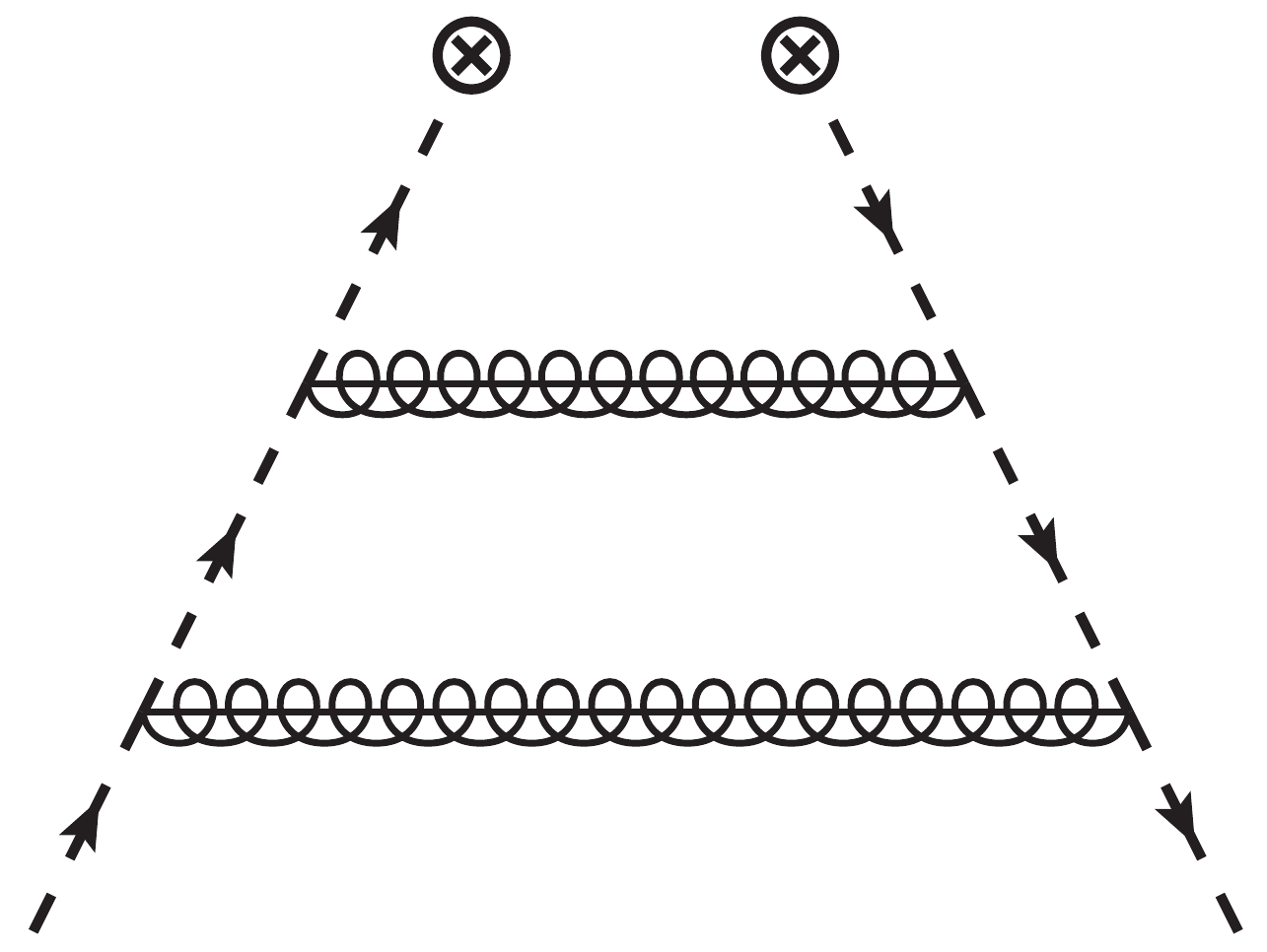}
\put(-100,68){a)}
\includegraphics[width=0.245\textwidth]{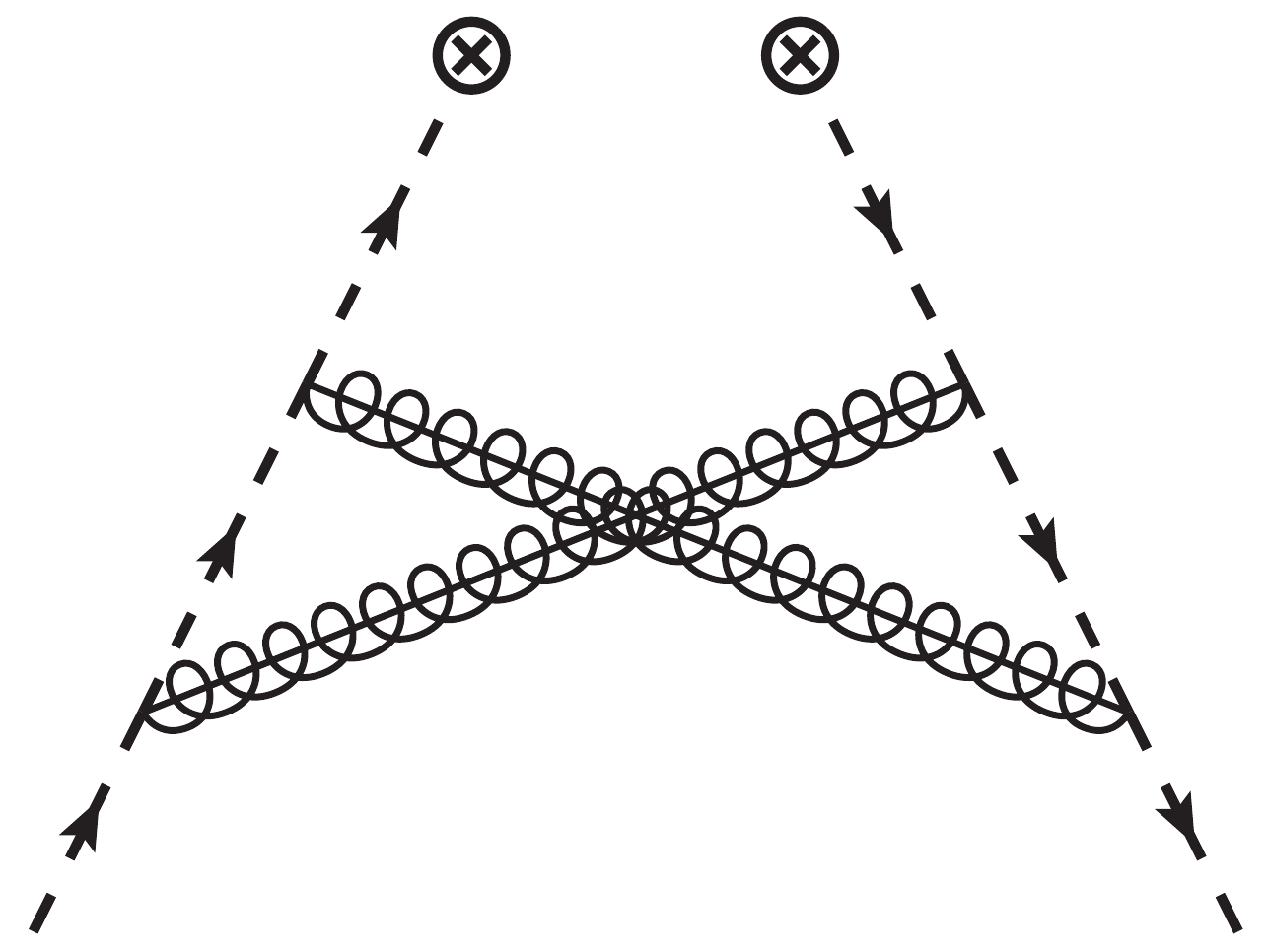}
\put(-100,68){b)}
\includegraphics[width=0.245\textwidth]{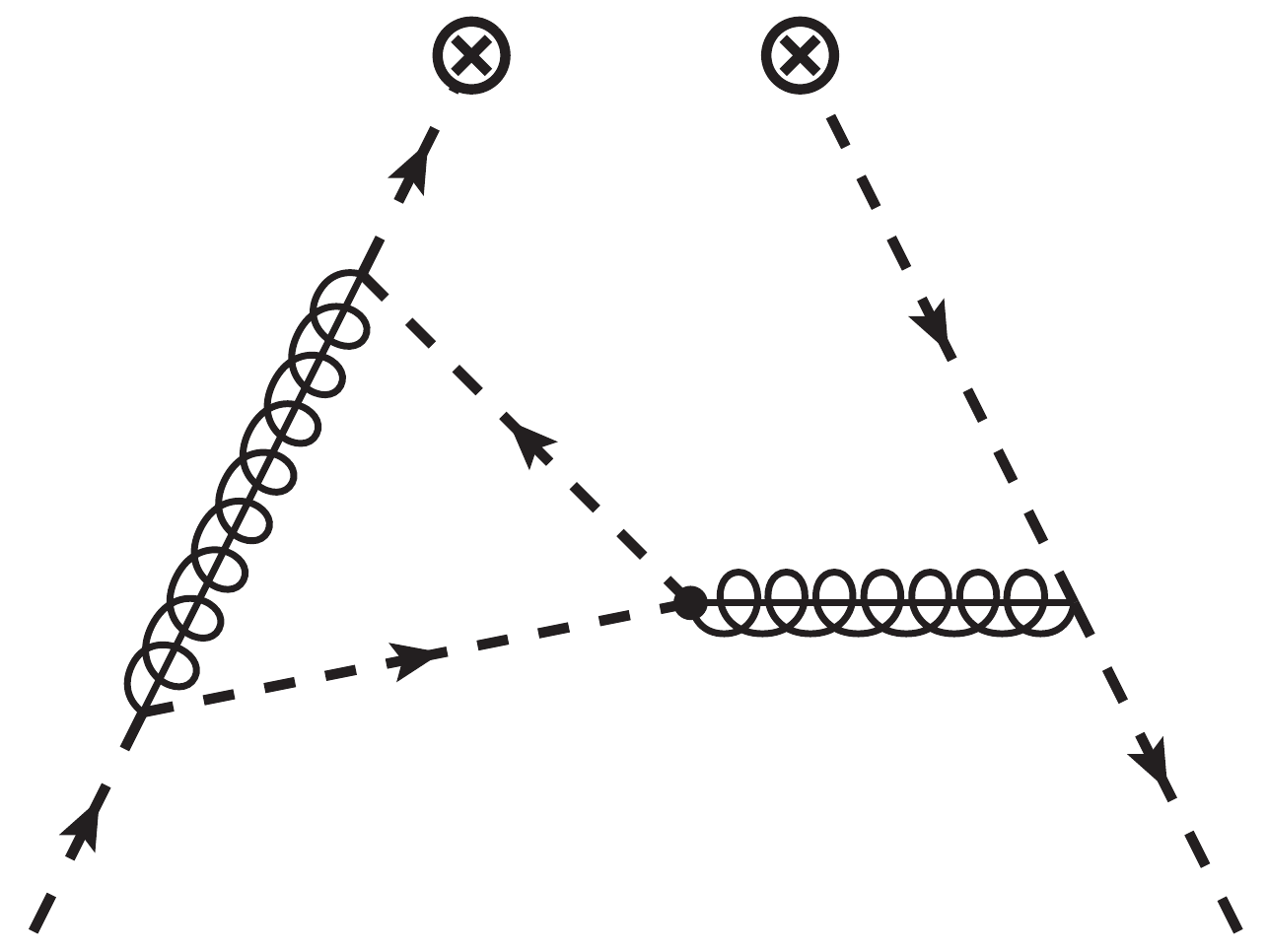}
\put(-100,68){c)}
\includegraphics[width=0.245\textwidth]{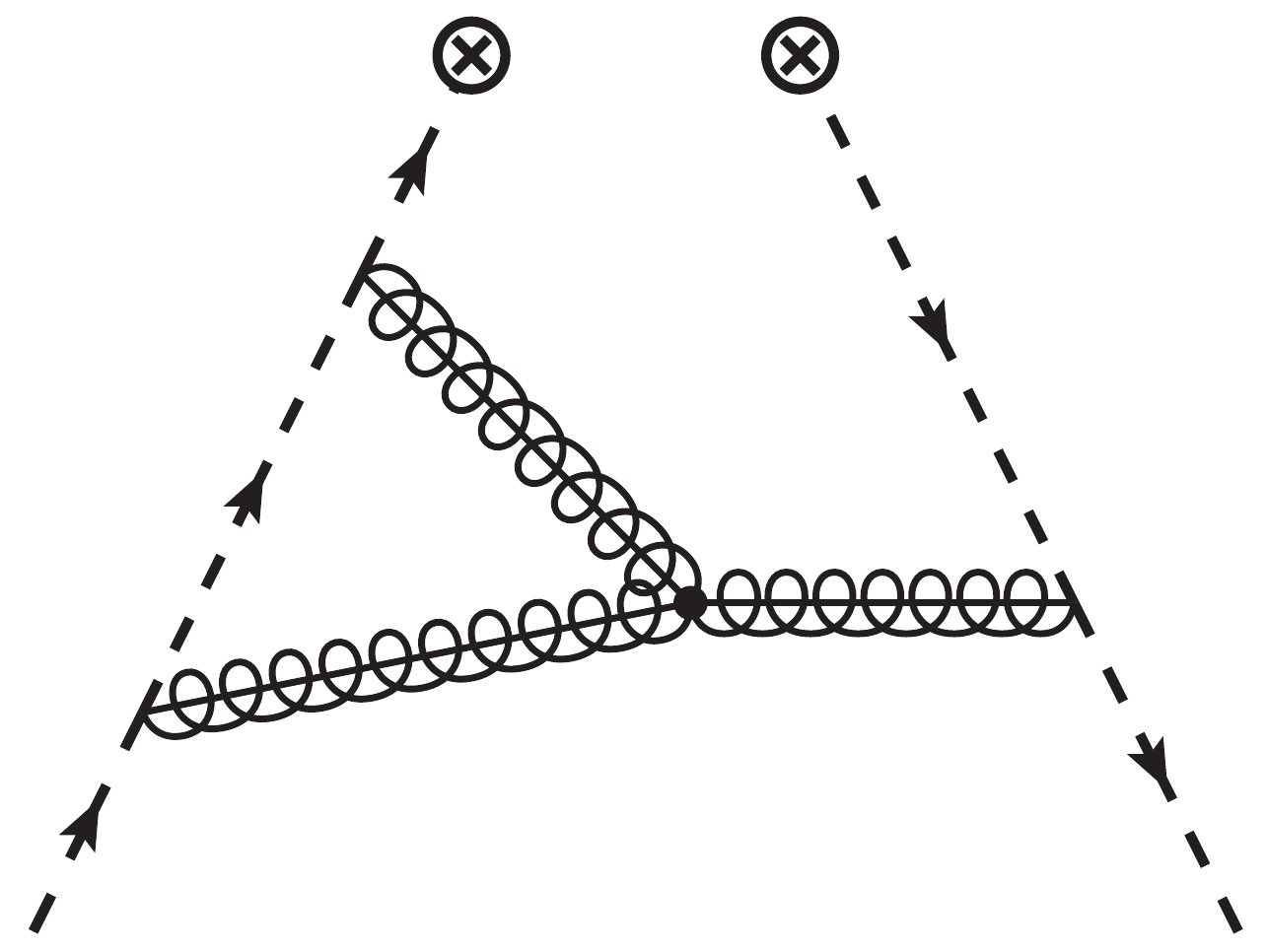}
\put(-100,68){d)}
\vspace{1ex}
\includegraphics[width=0.245\textwidth]{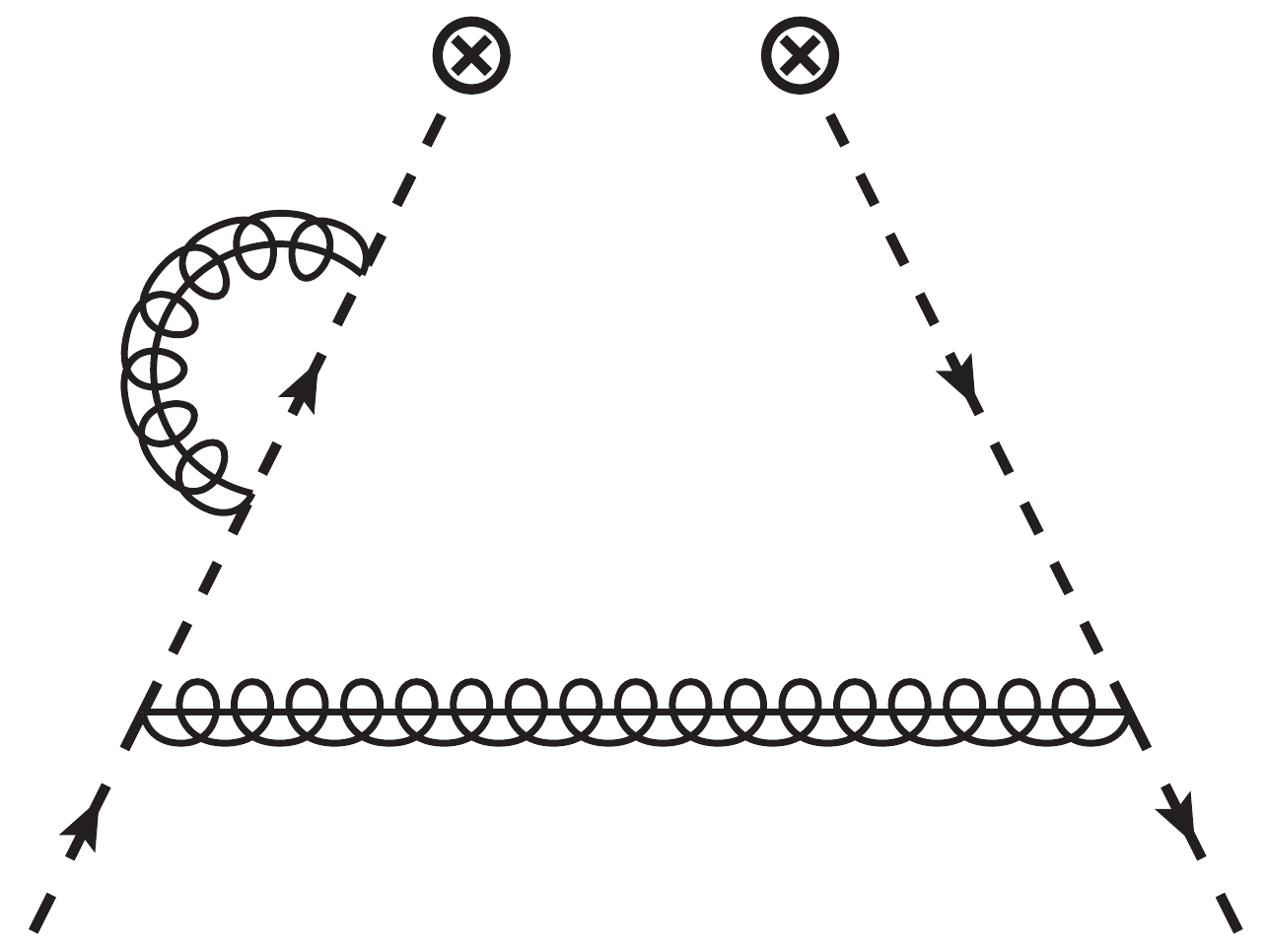}
\put(-100,68){e)}
\includegraphics[width=0.245\textwidth]{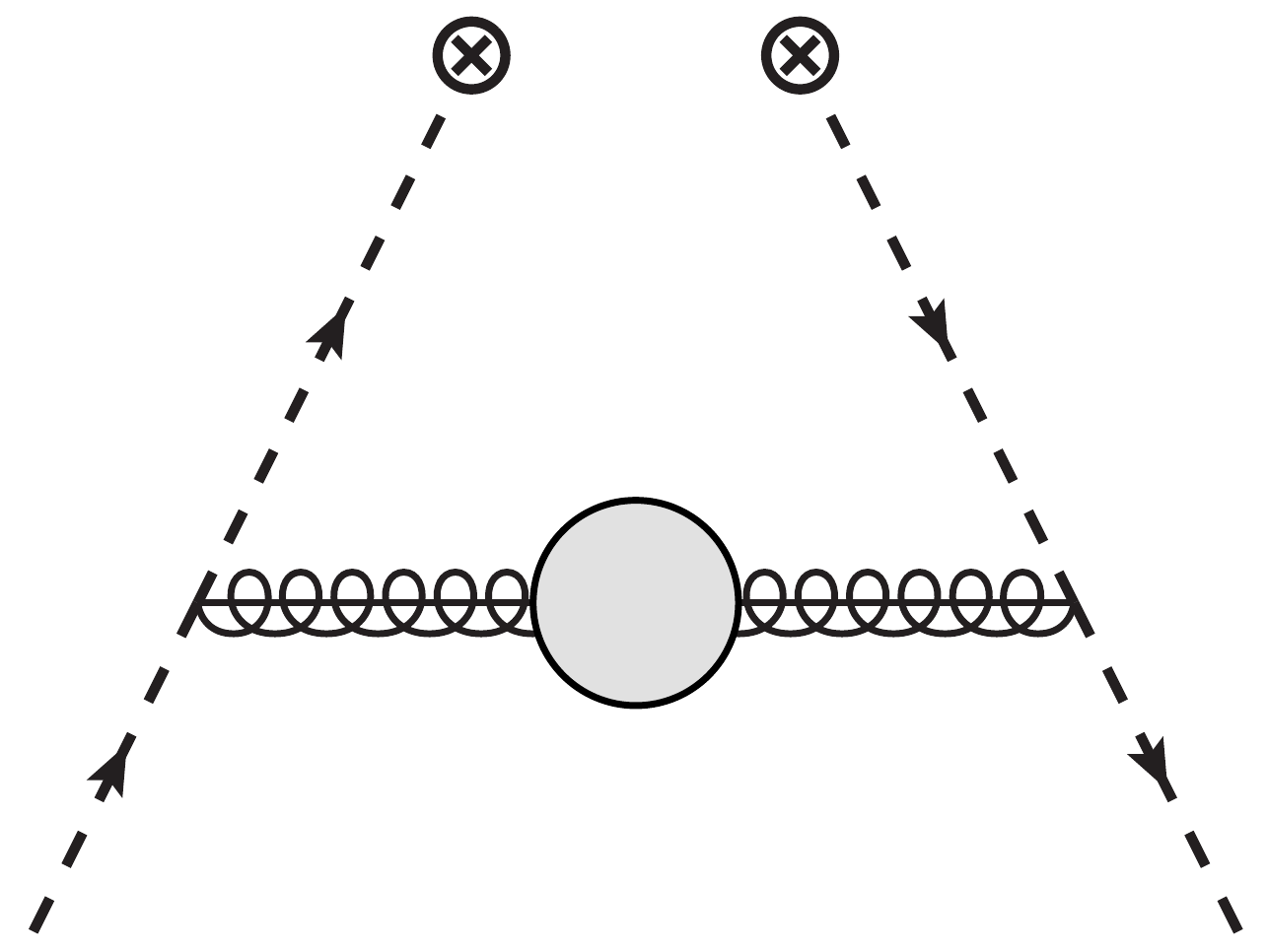}
\put(-100,68){f)}
\includegraphics[width=0.245\textwidth]{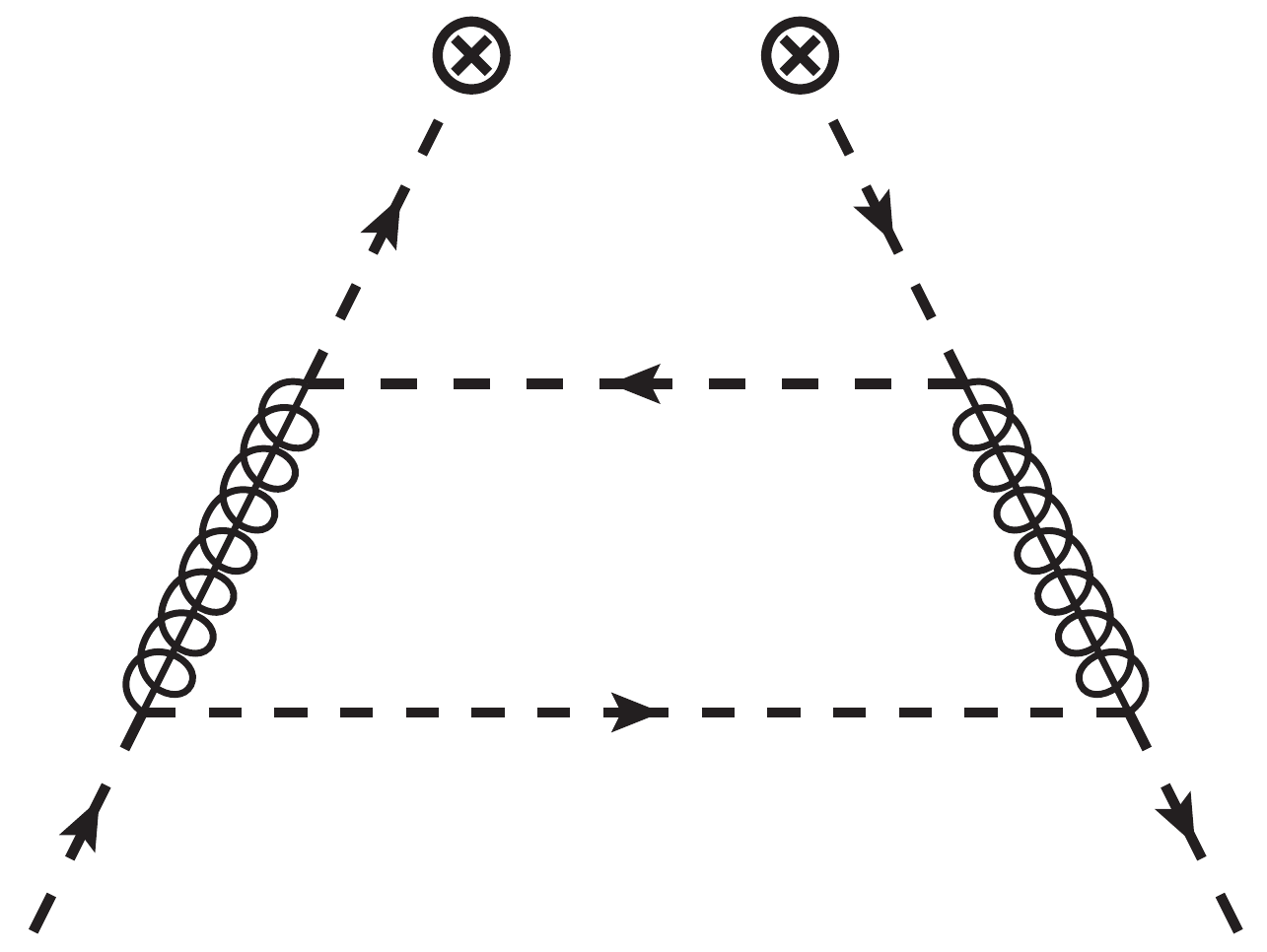}
\put(-100,68){g)}
\includegraphics[width=0.245\textwidth]{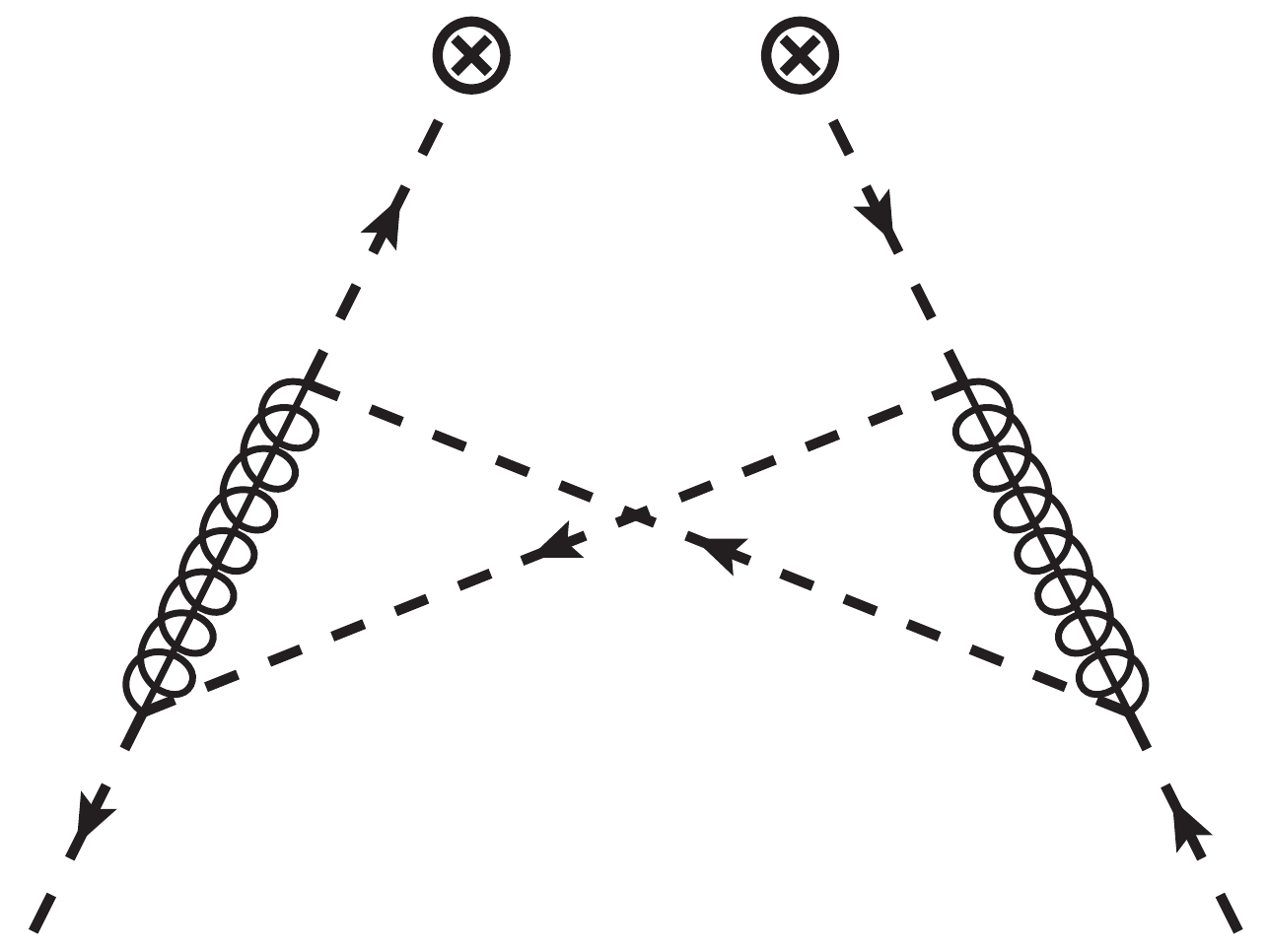}
\put(-100,68){h)}
\vspace{1ex}
\includegraphics[width=0.245\textwidth]{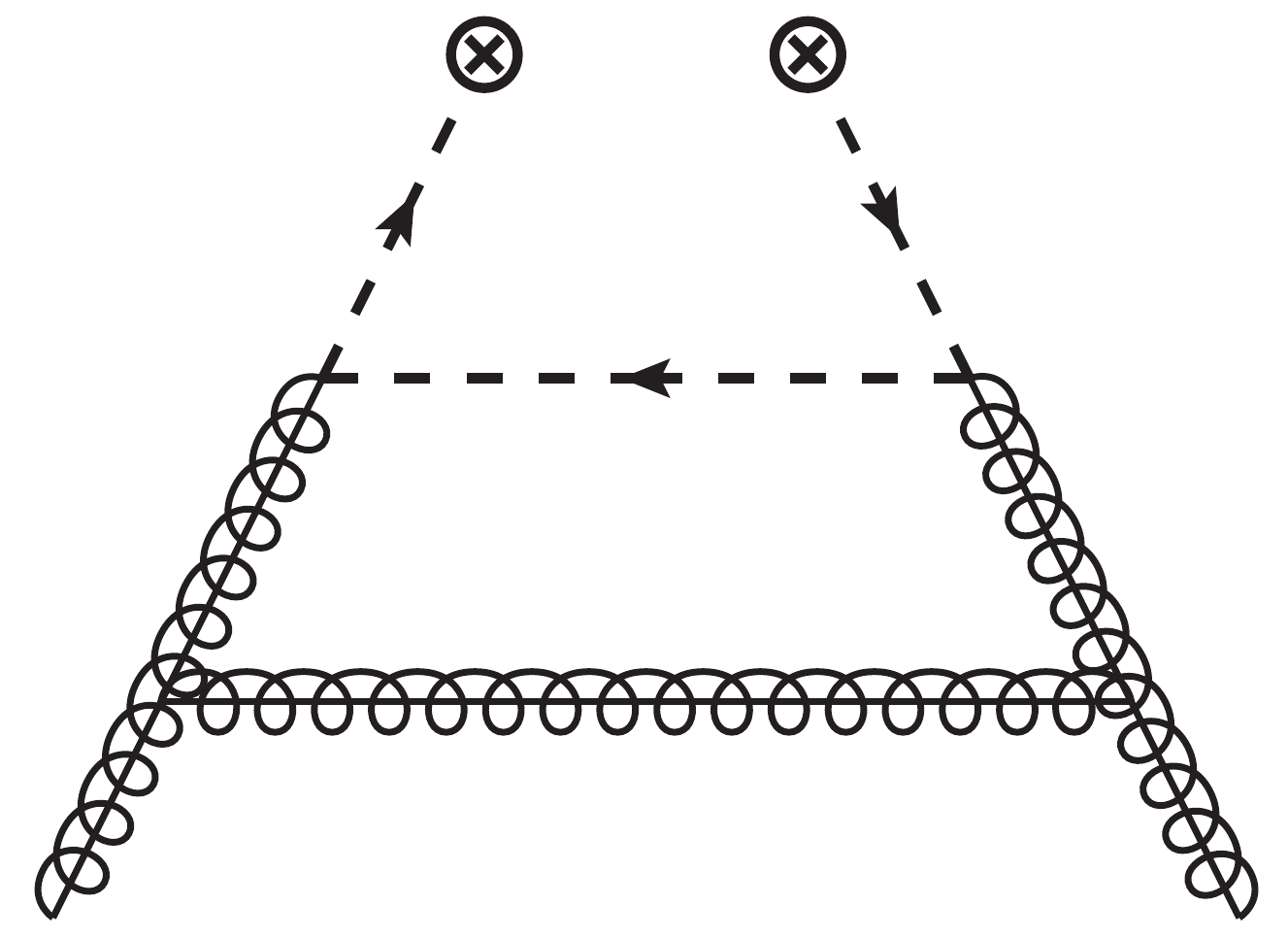}
\put(-100,68){i)}
\includegraphics[width=0.245\textwidth]{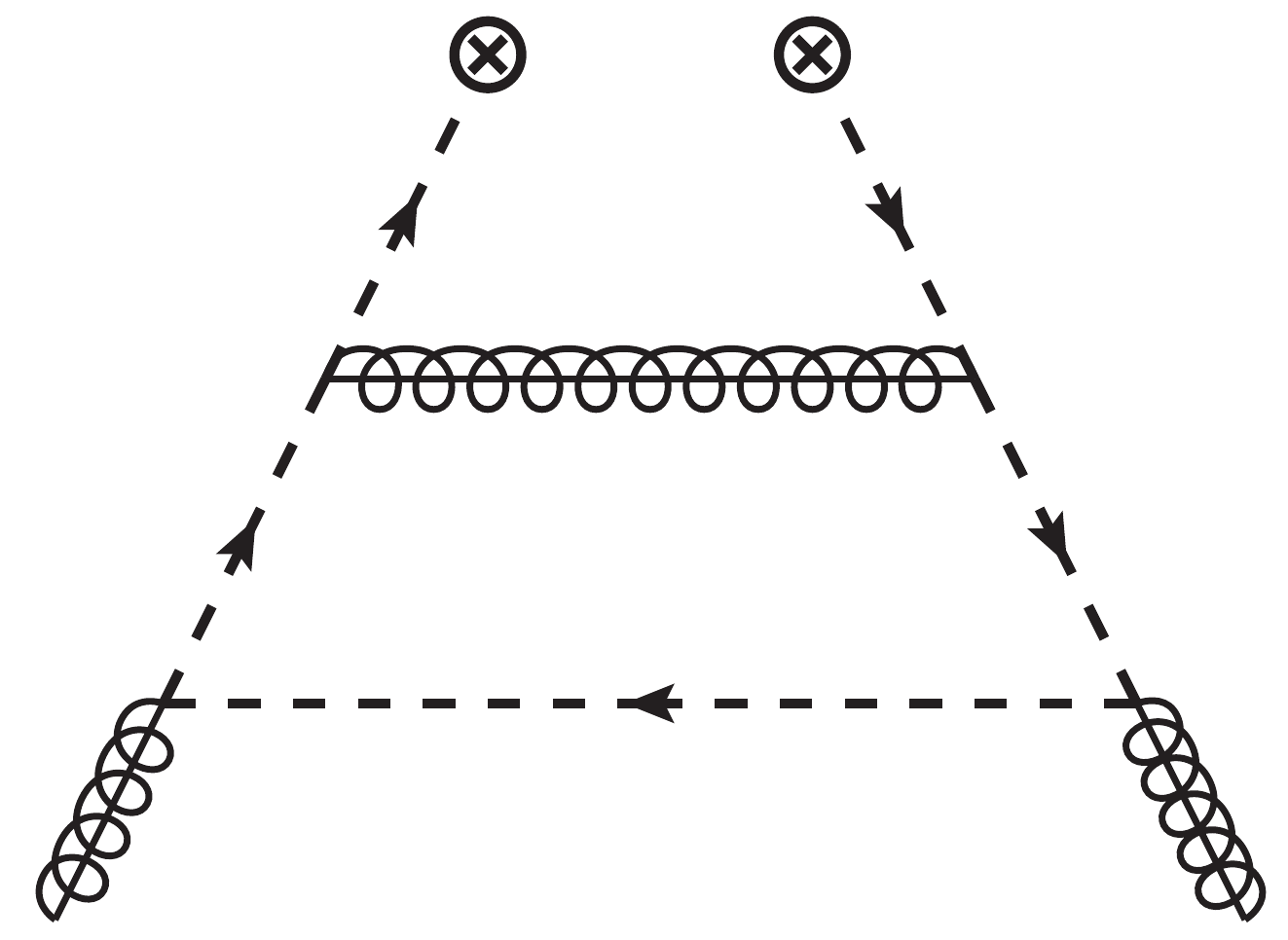}
\put(-100,68){j)}
\includegraphics[width=0.245\textwidth]{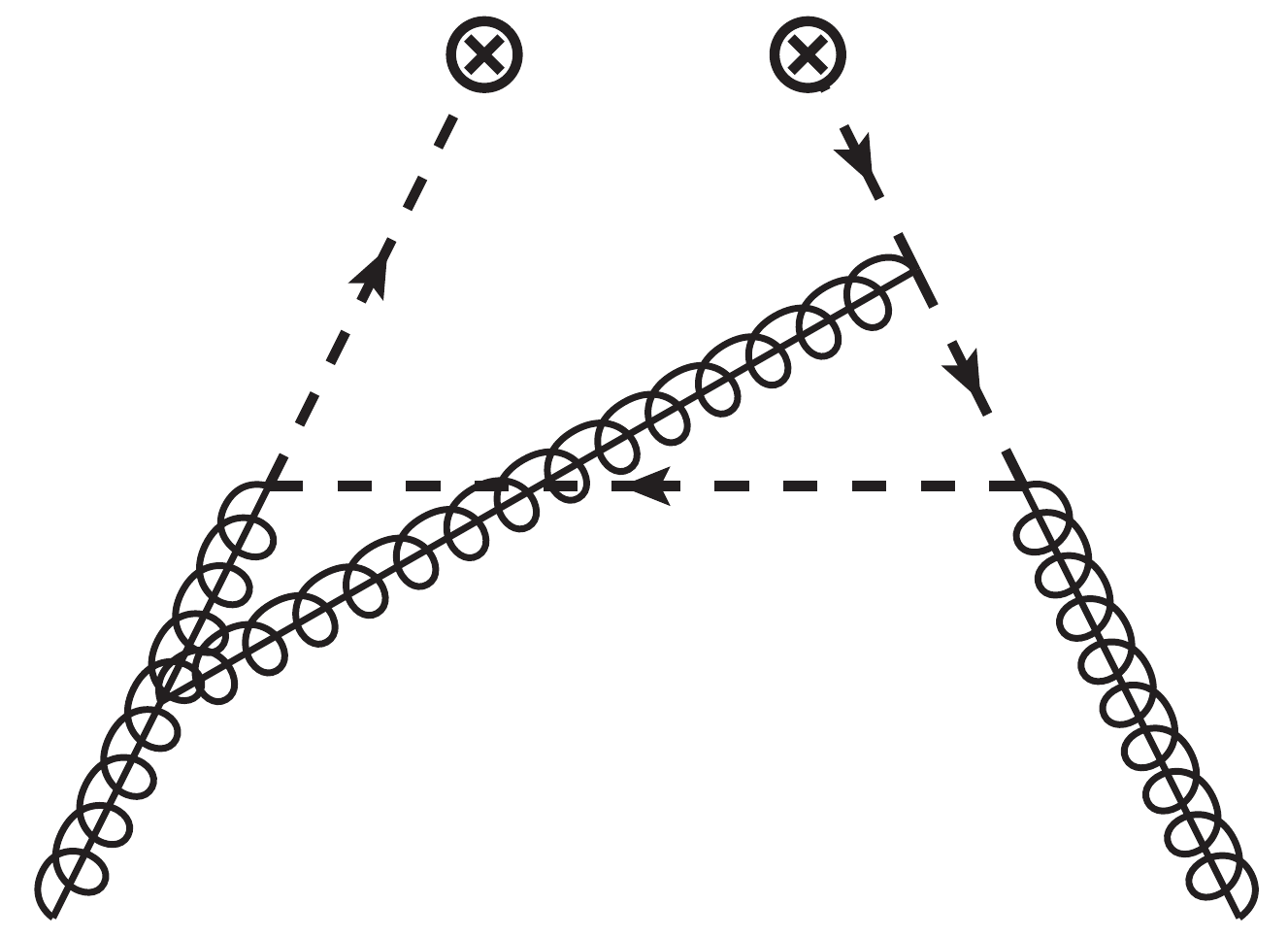}
\put(-100,68){k)}
\includegraphics[width=0.245\textwidth]{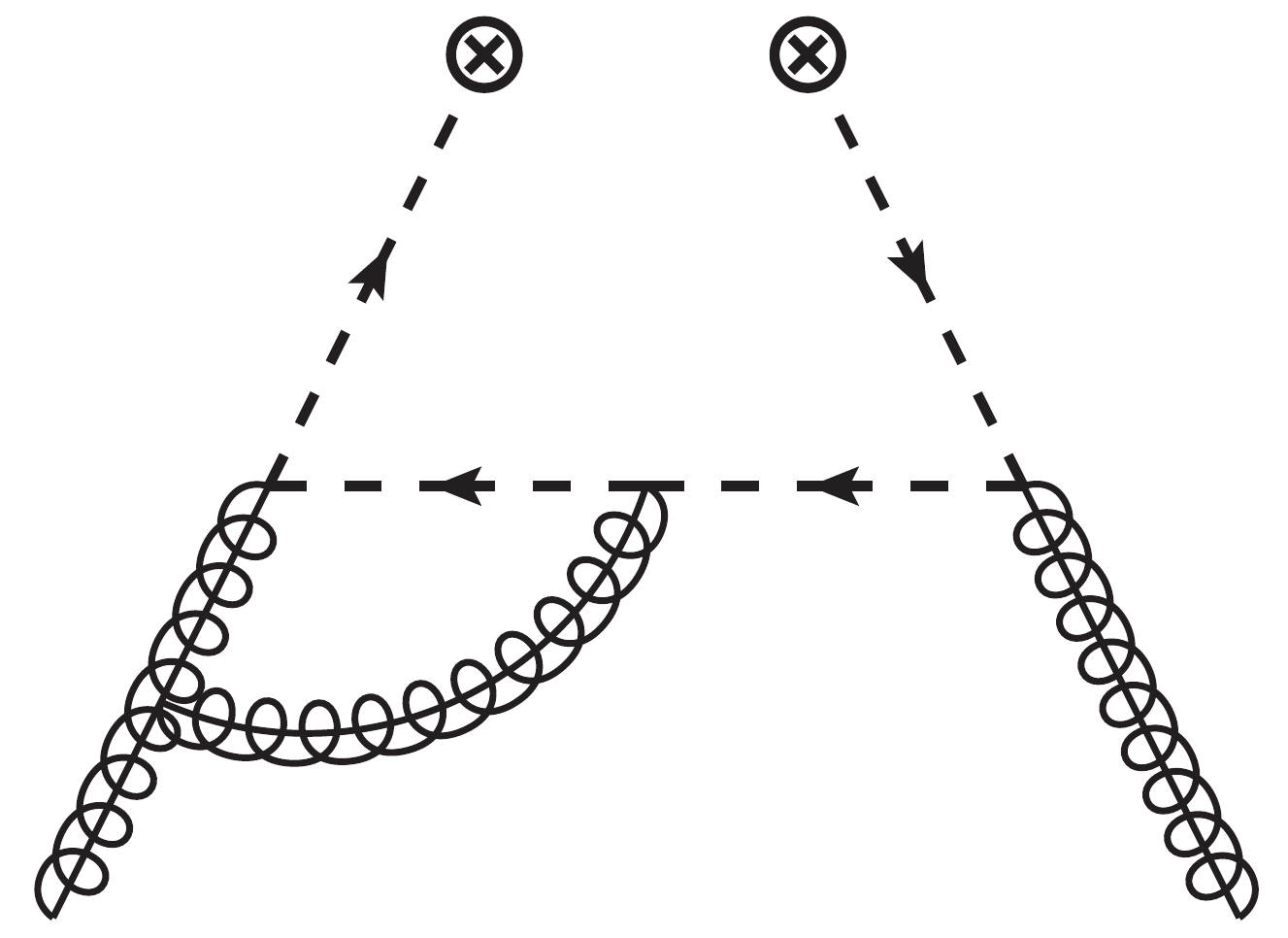}
\put(-100,68){l)}
\vspace{1ex}
\includegraphics[width=0.245\textwidth]{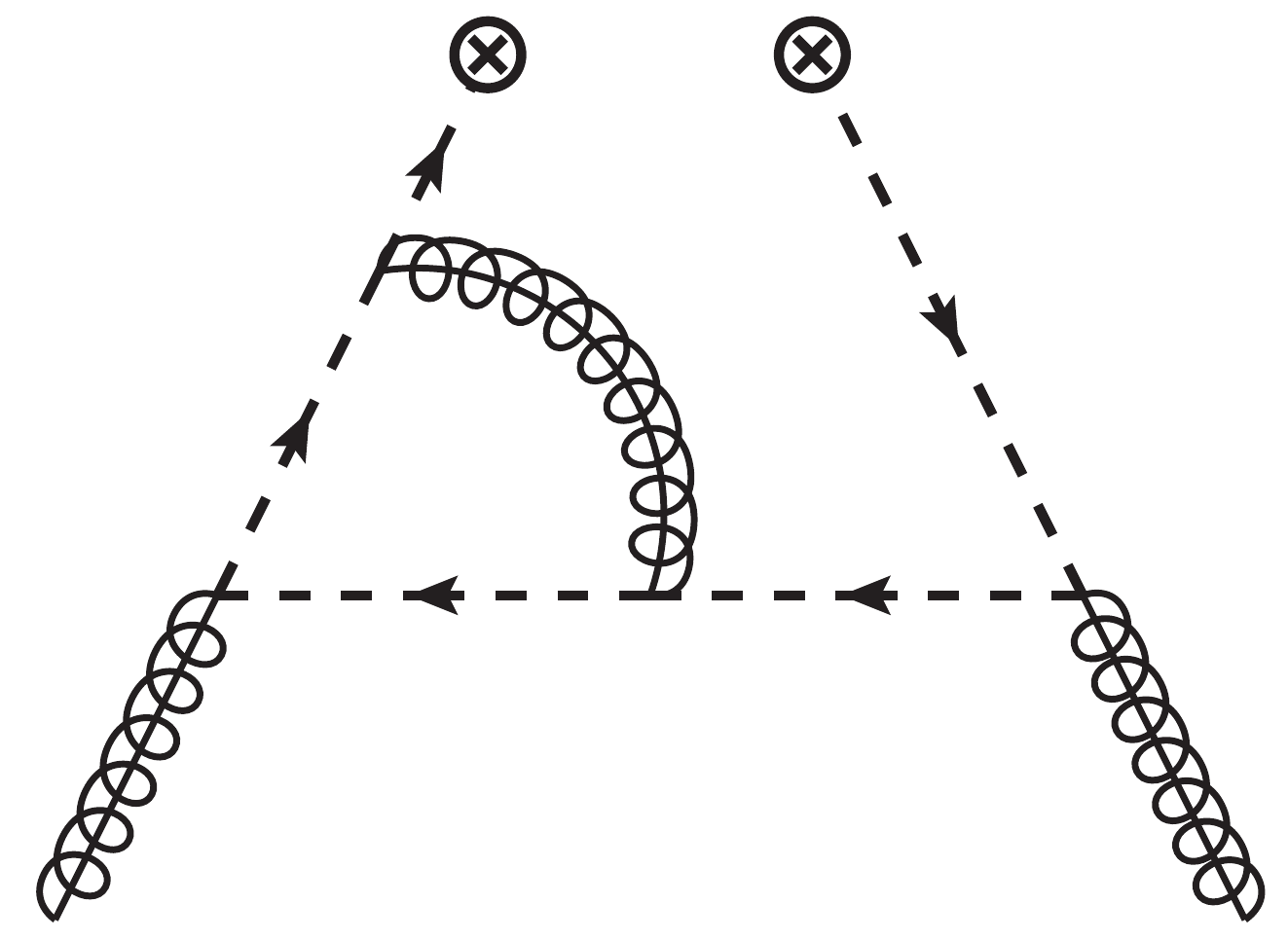}
\put(-100,68){m)}
\includegraphics[width=0.245\textwidth]{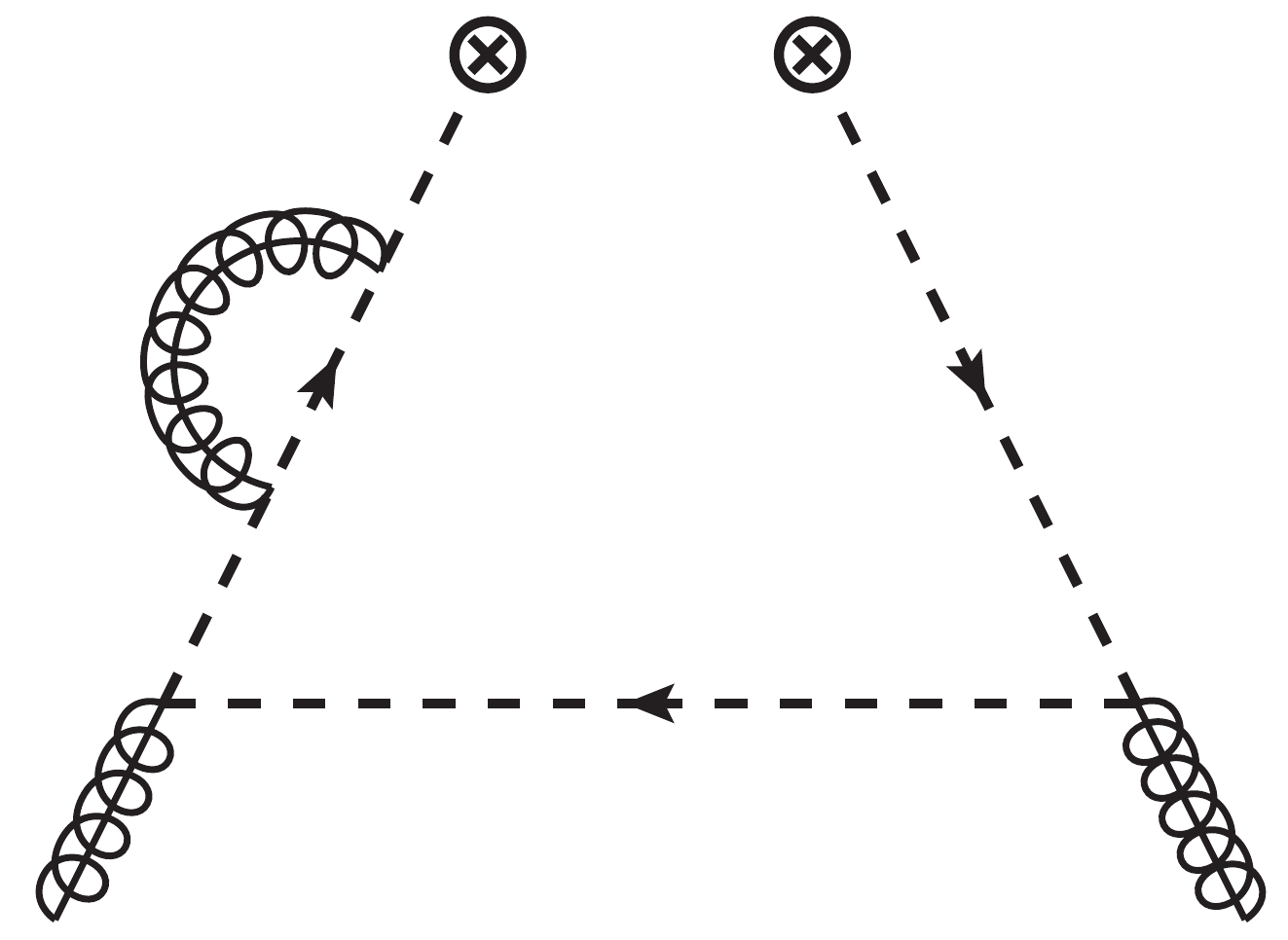}
\put(-100,68){n)}
\includegraphics[width=0.245\textwidth]{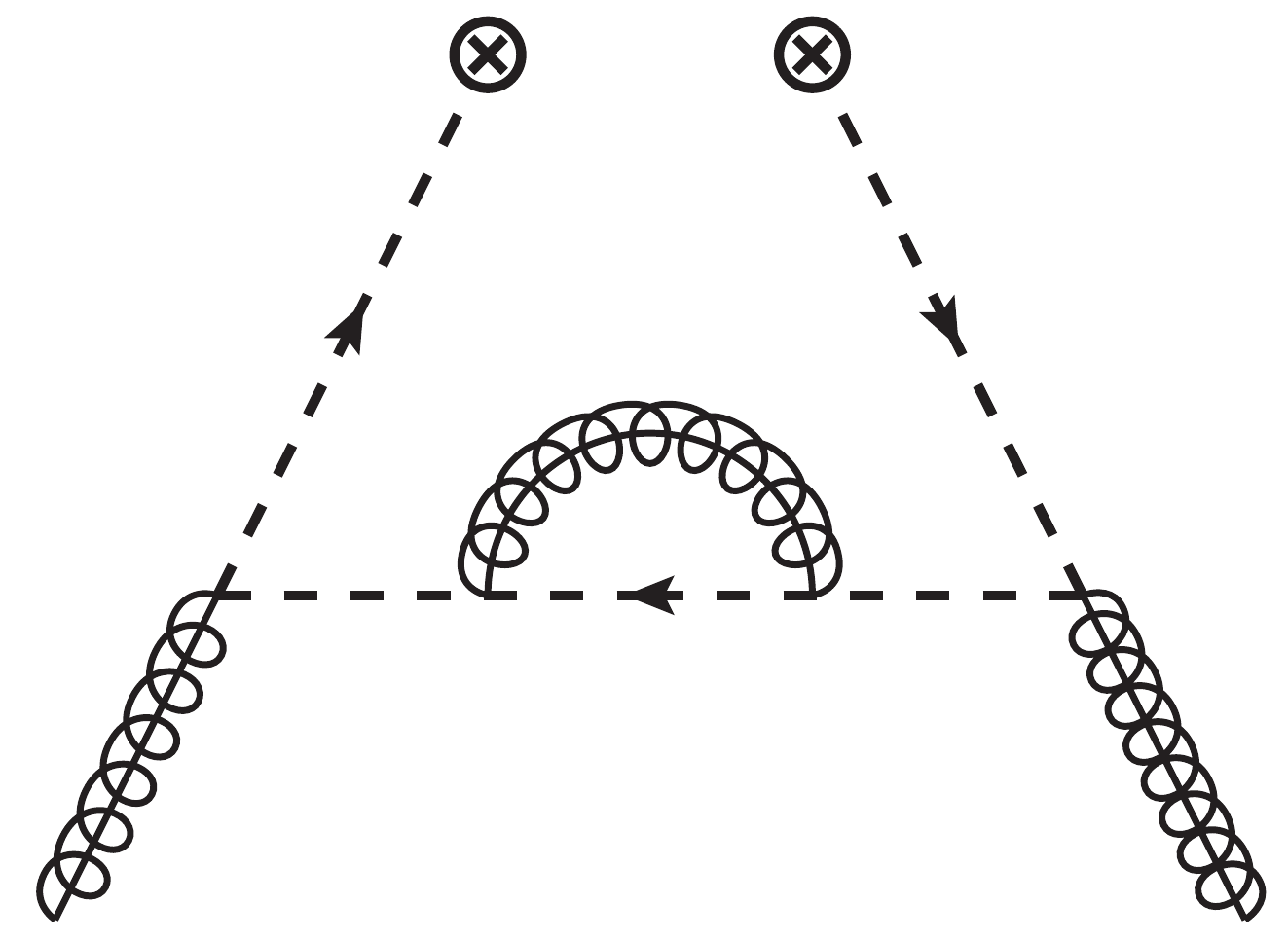}
\put(-100,68){o)}
\caption{
Diagrams contributing to the calculation of the NNLO matching coefficients $\cI_{q_iq_j}$ (a-g), $\cI_{q_i\bar{q}_j}$ (h) and $\cI_{q_ig}$ (i-o) using dimensional regularization. 
Left-right mirror graphs (up to the fermion flow) are not displayed. We also do not display the diagram analogous to~(g) but with the fermion flow of the external quarks reversed, which contributes to the flavor-singlet part of $\cI_{q_i\bar{q}_j}$.
The blob in diagram~(f) represents the full one-loop gluon self-energy. The graphs can either be interpreted as standard QCD diagrams or as SCET diagrams with collinear quark and gluon lines. Using axial gauge, this set of nontrivial diagrams is complete when using QCD Feynman rules, while it has to be supplemented by diagrams involving vertices of four collinear particles when using SCET Feynman rules. In Feynman gauge, additional diagrams with Wilson line connections as shown in \fig{WDiags} contribute.
\label{fig:BqDiagrams}}
\end{figure}

\begin{figure}[t]
\begin{center}
\includegraphics[width=0.245\textwidth]{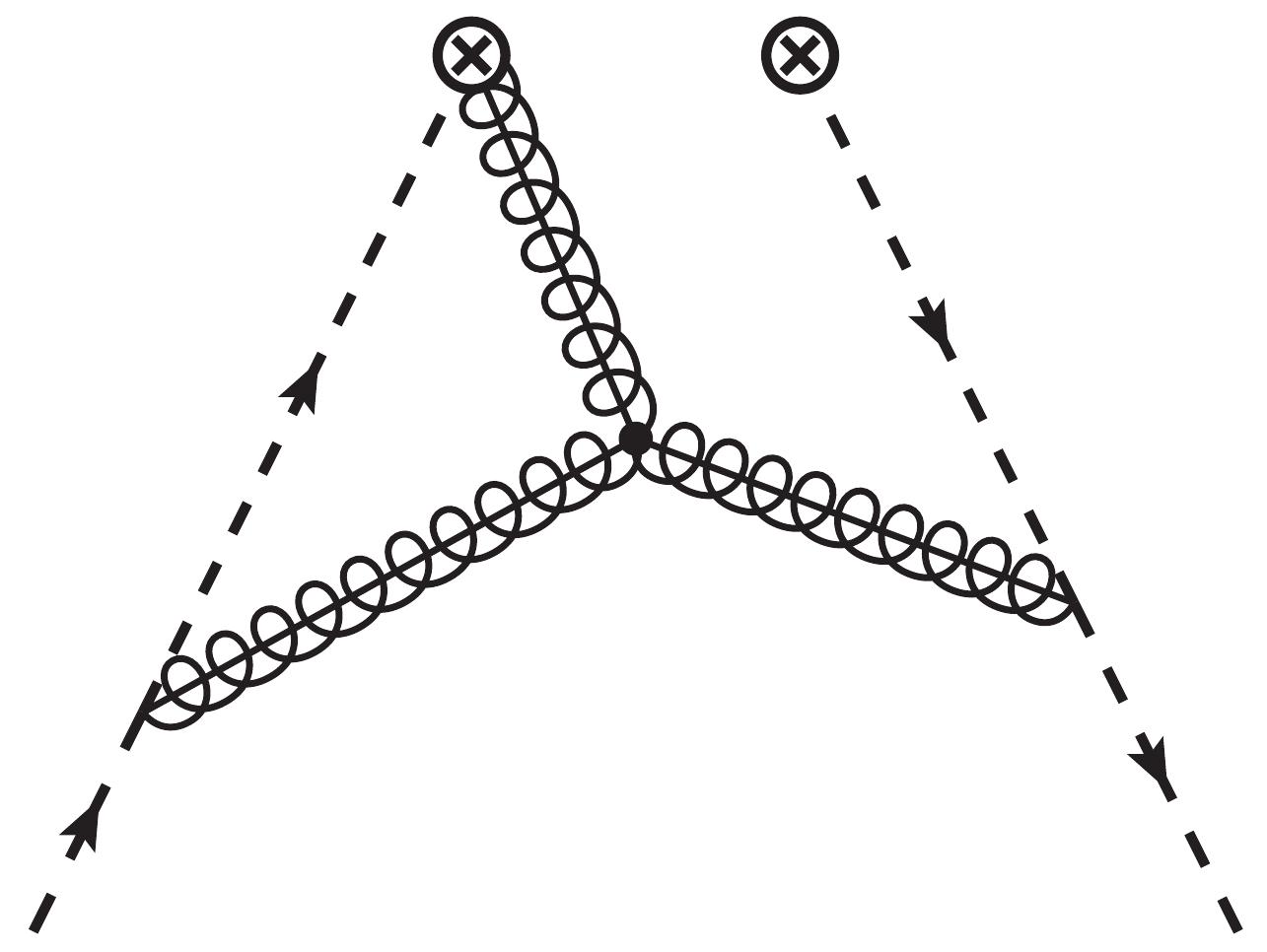}
\put(-100,68){a)}
\includegraphics[width=0.245\textwidth]{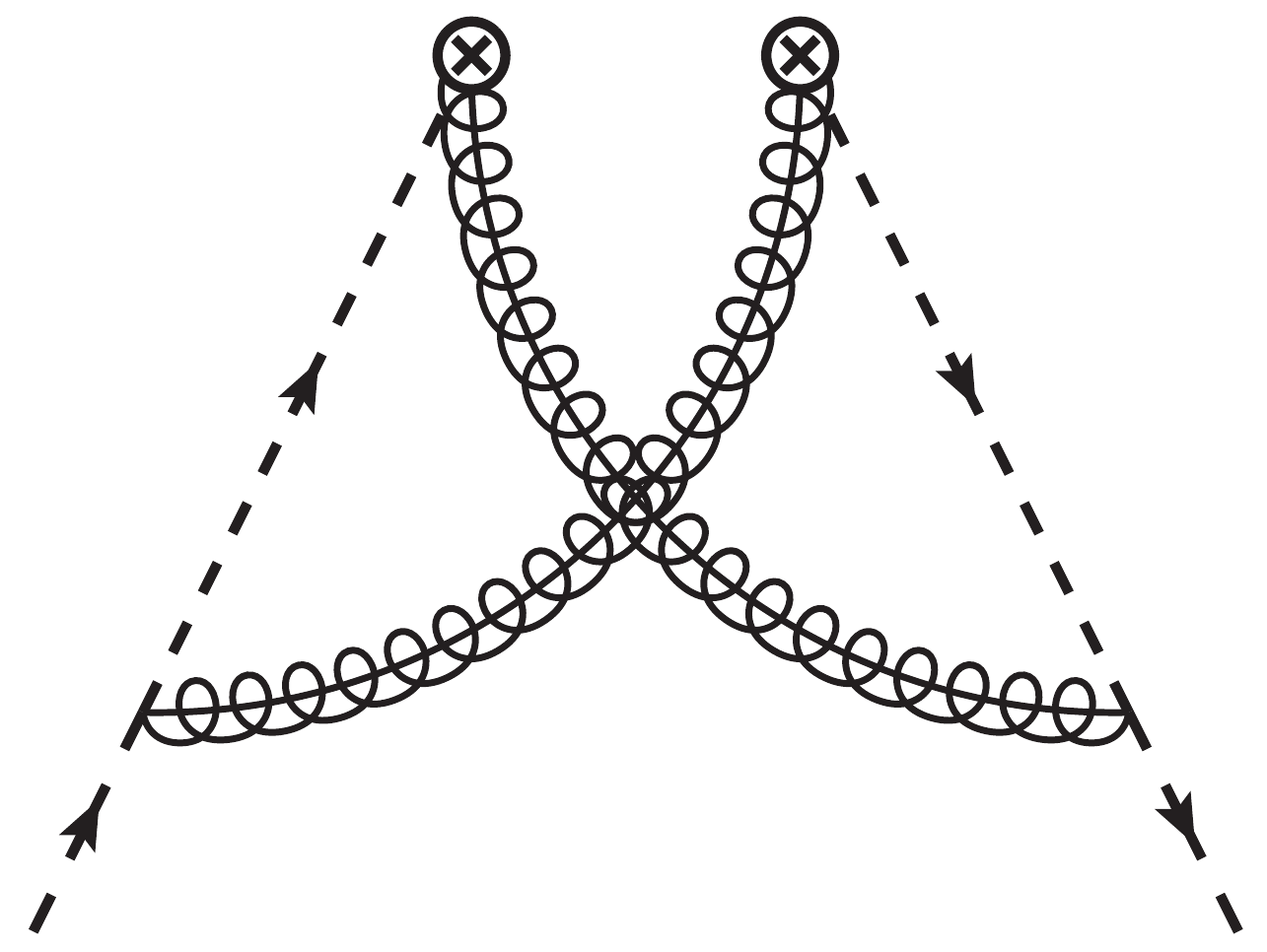}
\put(-100,68){b)}
\includegraphics[width=0.245\textwidth]{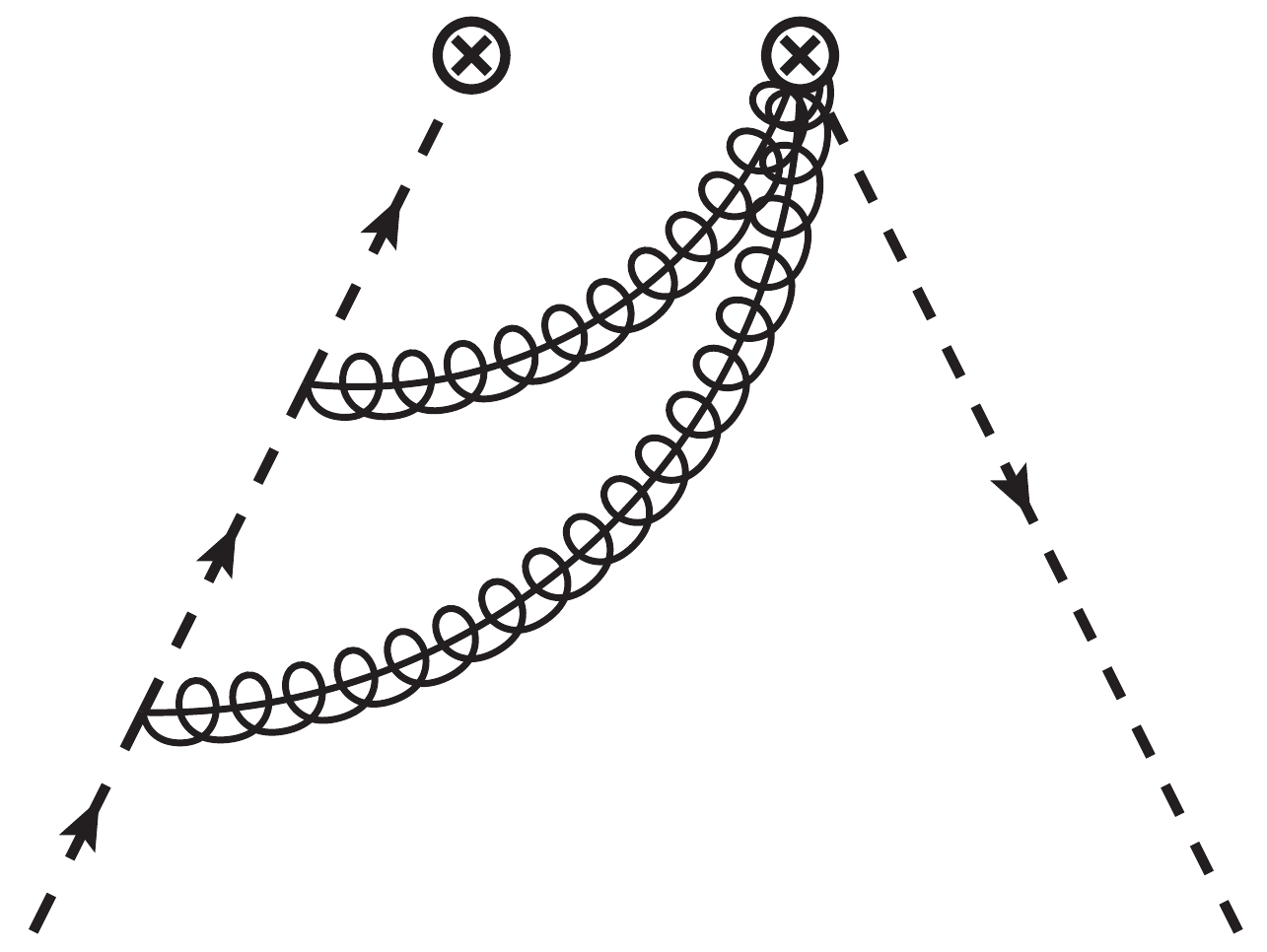}
\put(-100,68){c)}
\includegraphics[width=0.245\textwidth]{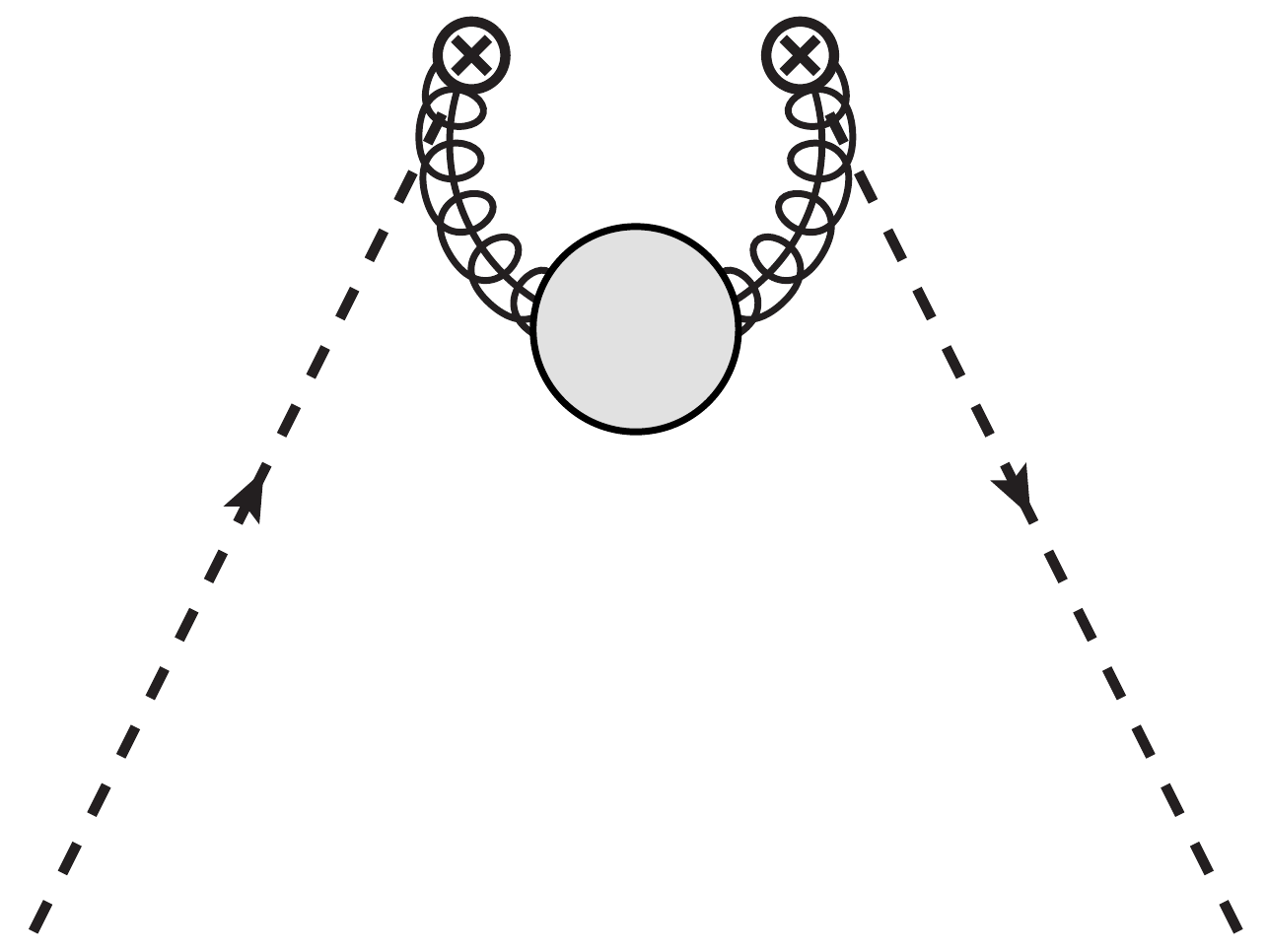}
\put(-100,68){d)}
\end{center}
\caption{Example diagrams with one (a) or two (b-d) connections to the collinear Wilson lines in the beam function operator [see \eqs{Wn}{BOpdefSCET}]. In Feynman gauge this class of diagrams has to be computed in addition to the diagrams shown in \fig{BqDiagrams}.
Diagram d) does not vanish, because the Lorentz structure of the gluon self energy insertion is nondiagonal in Feynman gauge.
\label{fig:WDiags}} 
\end{figure}

As explained in \mycite{Stewart:2010qs}, the bare partonic beam function matrix elements
may be calculated by computing the time-ordered partonic matrix elements
%%%
\begin{align} \label{eq:Tqdef}
\lefteqn{\Mae{j_n(p^-)}{\,T \{ \op^\bare_q(t, \omega) \}}{j_n(p^-)}} &  \\
&\qquad = \frac{\theta(\w)}{2\pi} \int\! \frac{\df y^-}{2\w}\, e^{\img (t/\w - p^+) y^-/2}
   \MAe{j_n(p^-)}{\,T\Bigl\{\bar\chi_n\Bigl(y^-\frac{n}{2}\Bigr) \frac{\bnslash}{2} \bigl[\delta(\w-\bnP_n) \chi_n(0)\bigr]\Bigr\}}{j_n(p^-)} \nn
%&\qquad=
%   \theta(\w) \,\MAe{j_n(p^-)}{\,T\Bigl\{ {\bar \chi_n(0)\, \delta(t - \w\hp^+)\, \frac{\bnslash}{2} \bigl[\delta(\w - \bnP_n) \chi_n(0)\bigr]} \Bigr\}}{j_n(p^-)}
\,,\end{align}
%%%
and taking their discontinuity [according to \eq{Disc_def2}]:
%%%
\begin{align} \label{eq:DiscTB}
B_{q/j}^\bare(t, \w) = \Disc_{t > 0}\, \Mae{j_n(p^-)}{\,T \{ \op^\bare_q(t, \omega) \}}{j_n(p^-)}
\,.\end{align}
%%%
We obtain the two-loop contributions to 
%$\Mae{j_n(p^-)}{T_{q}(t, \w)}{j_n(p^-)}$ 
\eq{Tqdef} from Feynman diagrams as shown in \fig{BqDiagrams}, where the insertion of the bilocal operator $T \{ \op^\bare_q(t, \omega) \}$ is denoted by two $\otimes$ symbols. In dimensional regularization all purely virtual diagrams are scaleless and vanish. The diagrams in \fig{BqDiagrams} are then the full set of relevant diagrams in the axial (light-cone) $\bar{n} \cdot A_{(n)} = 0$ gauge when using QCD Feynman rules. In this gauge the Wilson line operators in \eqs{chiSCET}{Wn} become equal to the identity. In contrast, in Feynman gauge one has to compute further diagrams, such as the ones shown in \fig{WDiags}, allowing for all possible connections of gluons to one or both of the Wilson lines. When using SCET Feynman rules, there are also additional diagrams involving the vertex of two collinear quarks and two collinear gluons.

%%%%%%

We evaluate the diagrams together with taking the discontinuity in \eq{DiscTB} using two different
methods, which we shall refer to as ``On-Shell Diagram Method'' and ``Dispersive Method''. They are described in \subsecs{cutmethod}{dispmethod} below.
When performing the calculation using the On-Shell Diagram Method, we use axial $\bar{n} \cdot A_{(n)} = 0$ gauge, while we use Feynman gauge in the calculation with the Dispersive Method. Both calculations yield the same result providing us with a strong cross check.\footnote{For $B^{\bare,(2)}_{q/q}$ we have in fact performed the Feynman gauge calculation using SCET Feynman rules and the axial gauge calculation using QCD Feynman rules as a further check. Otherwise we have always used QCD Feynman rules.}
We emphasize that the gauge choice is independent of the method and any gauge could have been used with either method.

In both methods the calculation of $B^{\bare,(2)}_{q/q}$, which is singular in the limit $z \to 1$ (just like $P_{qq}$), 
is divided into two stages. First, $B^{\bare,(2)}_{q/q}(z)$ is calculated for $z<1$, and then at the endpoint for $z\to 1$.
The advantage of this is that in the $z < 1$ calculation we can avoid one light-cone divergence, thereby avoiding having to expand master integrals to an extra order in
$\eps$. In the endpoint calculation, we can take the limit $z \to 1$ in an appropriate fashion, which
simplifies master integrals and allows us to more easily extend them to the extra order in $\epsilon$ required 
in this limit. Another possible way to calculate $B^{\bare,(2)}_{q/q}(z)$ near the endpoint is to replace the quark 
lines by eikonal (Wilson) lines in direction $n$ (following the direction of the quark), which then allows one to use 
web techniques~\cite{Gatheral:1983cz, Frenkel:1984pz} to calculate the graphs. In this way we checked the Abelian $C_F^2$ endpoint piece of the $qq$ beam function, which can be calculated entirely from the one-loop result given in \mycite{Stewart:2010qs} using the web technique.

The fact that the endpoint contribution of $B^{\bare,(2)}_{q/q}$ can be calculated by replacing the quark line by an
eikonal line can be used to make some definite statements about the form of the endpoint contributions, even without doing any
explicit calculations. The eikonal diagrams for the endpoint involve a Wilson line in the $n$ direction connected to
a Wilson line in the $\bar{n}$ direction on both sides of the cut, with all possible gluon connections between 
these. Four-momentum conservation requires that the summed momenta of the emitted gluons crossing the cut must have an $n$ component equal to $(1-z)P^-$. The $\bar{n}$ component is fixed by the measurement, cf. \eq{BOpdefSCET}, to be $t/(zP^-) \to t/P^-$ for $z \to 1$. Due to the symmetry between $n$ and $\bar{n}$ in the endpoint diagrams the result must be symmetric in $(1-z)P^-$ and $t/P^-$. Because $t$ is the only dimensionful Lorentz-invariant quantity involved, we can argue on dimensional grounds that $B^{\bare,(2)}_{q/q}$ must be proportional to $t^{-1-2\epsilon}$, since $t$ has mass dimension $2$ and at two loops we have a factor $\mu^{4\epsilon}$. Therefore, the overall endpoint contribution must be equal to $((1-z)t)^{-1-2\epsilon}$ multiplied only by a function of $\epsilon$. This result is useful to check our brute-force calculation of the endpoint contributions, which in individual diagrams gives terms like $(1-z)^{-1-\epsilon}t^{-1-2\epsilon}F(\epsilon)$ that must cancel between diagrams in the final result. Alternatively, it could be used to predict/check the $\delta(t) \mathcal{L}_n(1-z)$ for $n \neq 0$ terms in the
beam function from the $\delta(1-z) \mathcal{L}_n(t)$ terms for $n \neq 0$, whose form is in turn dictated by the anomalous
dimension $\gamma_B$ (see also \mycite{Procura:2011aq}).

Just as in the one-loop case~\cite{Stewart:2010qs}, the zero-bin subtractions in this two-loop calculation are 
scaleless and vanish. Their effect is to convert some $1/\epsilon^n$ divergences in $B^{\bare,(2)}_{q/j}$ from IR divergences to UV divergences, which are then renormalized by the counterterm $Z^q_B$ and contribute to the anomalous dimension $\gamma^q_B$.

%===============================================================================
\subsection{On-Shell Diagram Method}
\label{subsec:cutmethod}
%===============================================================================

This method employs the Cutkosky rules~\cite{Cutkosky:1960sp, Veltman:1994Ve} for computing the discontinuities,
and closely mirrors the method detailed in the paper by Ellis and Vogelsang~\cite{Ellis:1996nn}.
That is, each possible cut of the Feynman diagrams in \fig{BqDiagrams} is evaluated separately with
the particles crossing the cut being put on shell from the very beginning. The integral over the
phase space of the cut (real) partons as well as any integral over a virtual uncut loop momentum is performed for each diagram and each allowed cut.

The nonzero diagrams to be evaluated using this method can be divided into two classes: real-real diagrams in which two lines are cut and there are no loops either side of the cut, and real-virtual diagrams in which only one line is cut and there is a loop on one side of the cut. The real-virtual diagrams are computed with the aid of the integrals given in Appendix A of \mycite{Ellis:1996nn}, except converted to use dimensional regularization to regulate the IR divergences, and extended to one
higher order in $\epsilon$. For the virtual integral with a light-cone singularity in the center of the minus momentum integration
region, corresponding to eq.~(A.15) in \mycite{Ellis:1996nn}, it is
necessary to use a further regulator during the calculation, e.g. principal value, or modifying the power
of the light-cone divergence. The dependence on the regulator drops out in the final result.

The calculation of the real-real diagrams involves first using the on-shell constraints to perform the integrals
over plus momenta. Then the integrals over transverse momenta are done, with the delta function for $t$ being
used to fix the magnitude of one transverse momentum. Finally, the integrals over minus momenta are done, with
the other measurement delta function being used to fix one minus momentum. Just as in \mycite{Ellis:1996nn}, a suitable change
of transverse variables is performed to facilitate the integration over transverse momenta.

%===============================================================================
\subsection{Dispersive Method}
\label{subsec:dispmethod}
%===============================================================================

This method is a direct extension of the approach followed in \mycite{Stewart:2010qs} from one to two loops. Here, the matrix element in \eq{Tqdef} is obtained using standard perturbation theory in terms of loop diagrams like the ones shown in \figs{BqDiagrams}{WDiags} without cuts. The discontinuity is then taken after performing the integrations over most of the components of the loop momenta.

More precisely, we proceed as follows.
Letting $k$ and $l$ be the two loop momenta, two light-cone components, say $l^+$ and $l^-$, can be directly fixed by the (measurement) delta functions in the beam function operator, \eq{BOpdefSCET}.
The integration over the unconstrained $k^+$ momentum component we perform using residues. 
Nonzero contributions require poles in both halves of the complex $k^+$ plane, which consequently restricts the integration range for the $k^-$ momentum component to a finite interval.
After partial fraction decomposition of the resulting amplitude expressions we identify a set of master integrals. 
For the master integrals we introduce Feynman parameters to combine the propagator denominators and carry out the Euclidean transverse momentum integrations ($k_\perp$, $l_\perp$) in $d-2$ dimensions.
We now take the discontinuity according to \eq{disc_os2}. After that we perform the integration over the Feynman parameter(s) taking into account possible constraints on the integration range due to the $\theta$-function in \eq{disc_os2}.
Finally we integrate the amplitudes over $k^-$ and express the results as a function of $z=\w/p^-=l^-/p^-$.
Depending on the complexity of the master integral, it is convenient to expand the integrand in $\eps$ (to high enough order) before or after the Feynman parameter integration. In either case, we use the distributional identity in \eq{plus_exp} to consistently treat the $1/\eps^n$ poles induced by the Feynman parameter and/or $k^-$ integrations.

After the $\eps$ expansion the $k^-$ integration is finite for most of the two-loop diagrams. Feynman gauge diagrams with a triple gluon vertex and connections to Wilson lines, like the one in \fig{WDiags}~a, however, exhibit light-cone singularities within the $k^-$ integration region. As in the On-Shell Diagram method these divergences require further regularization. Since in the Dispersive Method the analogy to the two-loop PDF calculation in \mycite{Ellis:1996nn} is somewhat obscured and the distinction between real-real and real-virtual contributions is often not clear we find it most straightforward to modify the power of the light-cone divergence in order to regulate it. Namely, we chose the $\eta$-regulator proposed in \mycite{Chiu:2012ir}, which assigns a noninteger $1+\eta$ power to each Wilson line propagator. After the $k^-$ integration, the divergent $1/\eta$ terms cancel for each two-loop diagram individually and we can safely take the $\eta\to 0$ limit to obtain a regulator-independent result.

In contrast to the On-Shell Diagram Method, the viability of the Dispersive Method does not rely on a suitable choice of integration variables. On the other hand, for each diagram only the sum of all possible cuts can be determined directly and it is hard (if not impossible) to disentangle the contributions from individual cuts, which are the natural outcome of the On-Shell Diagram Method. This prohibits separate cross checks within individual classes of real-real and real-virtual diagrams at intermediate steps of the calculation. Overall, the calculational effort in order to determine $B^{\bare,(2)}_{q/j}$ is similar for both methods. For more details on technical issues in the two-loop beam function calculation we refer to an upcoming publication~\cite{Gaunt:2014cfa}.

%%%%%%%%%%%%%%%%%%%%%%%%%%%%%%%%%%%%%%%%%%%%%%%%%%%%%%%%%%%%%%%%%%%%%%%%%%%%%%%%
\section{Results}
\label{sec:results}
%%%%%%%%%%%%%%%%%%%%%%%%%%%%%%%%%%%%%%%%%%%%%%%%%%%%%%%%%%%%%%%%%%%%%%%%%%%%%%%%

Here we present our new results for the two-loop quark matching functions $I^{(2)}_{ij}(z)$ entering in \eq{I2master} for $i = q$ and $i = \bar q$. Since QCD is charge conjugation invariant we can easily obtain the antiquark coefficients $I^\two_{\bar{q}j}(z)$ from our results for $I^\two_{qj}(z)$. We decompose the two-loop coefficients as follows
%%%
\begin{align}
I_{\bar q_i \bar q_j}^\two(z) = I_{q_i q_j}^\two(z)
&= C_F\, \theta(z) \bigl[ \delta_{ij} I_{qqV}^\two(z) + I_{qqS}^\two(z) \bigr]
\,, \nn \\
I_{\bar q_i q_j}^\two(z) = I_{q_i \bar q_j}^\two(z)
&= C_F\, \theta(z) \bigl[ \delta_{ij} I_{q\bar qV}^\two(z) + I_{qqS}^\two(z) \bigr]
\,, \nn \\
I_{\bar q_i g}^\two(z) = I_{q_i g}^\two(z)
&= T_F\, \theta(z) I_{qg}^\two(z)
\,,\end{align}
%%%
where $q_i$ ($\bar q_i$) denotes the (anti)quark of flavor $i$ and find the following results:
%%%
\begin{align} \label{eq:IqqV_two}
I_{qqV}^\two(z)
&= \delta(1-z) \Bigl[ C_F \frac{7 \pi^4}{120} + C_A \Bigl( \frac{52}{27} - \frac{\pi^2}{6} - \frac{\pi^4}{36} \Bigr)
      + \beta_0 \Bigl( \frac{41}{27} - \frac{5 \pi^2}{24} - \frac{5 \zeta_3}{6} \Bigr) \Bigr]
   \nn \\ & \quad
   + C_F \biggl\{
      (1+z^2) \Bigl[\cL_3(1-z) - \frac{5\pi^2}{6} \cL_1(1-z) + 4 \zeta_3 \cL_0(1-z) \Bigr]
      - \frac{2}{1-z}\, T_3(z)
      \nn \\ & \qquad
      + \frac{1 + z^2}{1 - z} \bigl[V_3(z) - 2 U_3(z) \bigr]
      + \Bigl(\frac{1}{1 - z} + z \Bigr) \Bigl(3 \Li_2(z) + \frac{9}{4} \ln^2z + 4 \ln z- \frac{\pi^2}{2} \Bigr)
      \nn \\ & \qquad
      - (1 - z) \Bigl[\ln(1-z)\ln z - \frac{\pi^2}{6}\Bigr]
      + \Bigl(-\frac{1}{2} + \frac{3z}{4} \Bigr) \ln^2z
      + \Bigl(-6 + \frac{11 z}{2} \Bigr) \ln(1-z)
      \nn \\ & \qquad
      - \Bigl(\frac{5}{2} + 18 z \Bigr) \ln z
      - \frac{13 - 15 z}{2}
         \biggr\} 
   \nn \\ & \quad
   + C_A \biggl\{
      (1+z^2) \Bigl[ \Bigl(\frac{2}{3} - \frac{\pi^2}{6}\Bigr) \cL_1(1-z) + \Bigl(-\frac{8}{9} + \frac{7 \zeta_3}{2}\Bigr)\cL_0(1-z) \Bigl]
      \nn \\ & \qquad
      + \frac{1 + z^2}{1 - z} \Bigl[ U_3(z) - \ln(1-z)\ln\frac{1 - z}{z} \ln z - \frac{5}{4} \ln z \Bigr]
      - (1 + z) \Li_2(z)
      - \frac{3z}{2} \ln^2z
      \nn \\ & \qquad
      + \Bigl(3 - \frac{5 z}{2} \Bigl) \ln(1-z)
      - \frac{1 - 11 z}{2} \ln z
      + \frac{7 - 11 z}{4}
      + (1 + 3 z) \frac{\pi^2}{12}
   \biggr\}
   \nn \\ & \quad
   + \beta_0 \biggl\{
      (1+z^2)\Bigl[-\frac{1}{4} \cL_2(1-z) + \frac{5}{6} \cL_1(1-z) + \Bigl(-\frac{7}{9} + \frac{\pi^2}{12}\Bigr) \cL_0(1-z) \Bigr] 
      \nn \\ & \qquad
      + \frac{1 + z^2}{1 - z} \Bigl[ \frac{1}{2}\Li_2(1-z) + \ln(1-z)\ln z - \frac{5}{8} \ln^2 z - \frac{5}{4} \ln z \Bigr]
      \nn \\ & \qquad
      + \frac{1 - z}{2} \Bigl[\ln(1-z) + \frac{1}{2} \Bigr]
      + \frac{z}{2} \ln z
   \biggr\}
\,,\end{align}
%%%
%%%
\begin{align} \label{eq:IqqbarV_two}
I_{q\bar qV}^\two(z)
&= (2 C_F - C_A) \biggl\{ \frac{1 + z^2}{1+z}\, S_3(z)
   - \frac{1 + z}{2} \Bigl[ \Li_2(z^2) + 2 \ln z \ln(1+z) - \frac{\pi^2}{6} \Bigr]
   - z \ln^2 z
   \nn \\ & \qquad
   + 2 (1 - z) \ln(1-z)
   + \frac{3 + 19 z}{4} \ln z
   + \frac{15}{4} (1 - z)
   \biggr\}
\,,\end{align}
%%%
%%%
\begin{align} \label{eq:IqqS_two}
I_{qqS}^\two(z)
&= T_F \biggl\{ -2 (1 + z)\, T_3(z)
   - \Bigl(3 + \frac{4}{3 z} + 5 z + \frac{8 z^2}{3}\Bigr) \Bigr[\Li_2(z) - \frac{\pi^2}{6}\Bigr]
   \nn \\ & \qquad
   + \Bigl(1 + \frac{4}{3 z} - z - \frac{4 z^2}{3}\Bigr) \Bigl[\frac{1}{2} \ln^2(1-z)
   - \ln(1-z)  \ln z - \frac{\pi^2}{6} \Bigr]
   \nn \\ & \qquad
   - \Bigl[\frac{13}{4} (1 + z) + \frac{10 z^2}{3}\Bigr] \ln^2 z
   + \Bigl(\frac{26}{9z} - \frac{11}{3} + \frac{17z}{3} - \frac{44 z^2}{9} \Bigr) \ln(1-z)
   \nn \\ & \qquad
   +  \Bigl(\frac{23}{3} - \frac{5 z}{3} + \frac{76z^2}{9} \Bigr) \ln z
   +  \frac{104}{27z} - \frac{41}{18} + \frac{17 z}{18} - \frac{68 z^2}{27} 
   \biggr\}
\,,\end{align}
%%%
\begin{align} \label{eq:Iqg_two}
I_{qg}^\two(z)
&= C_F \biggl\{
      - 2 (1 - z)^2\, T_3(z)
      + P_{qg}(z) \Bigl[V_3(z) + \frac{5}{6} \ln^3(1-z) - \ln\frac{1-z}{z} \ln(1-z) \ln z
      \nn \\ & \qquad
         - \frac{5\pi^2}{6} \ln(1-z) - \frac{\pi^2}{3} \ln z + 11 \zeta_3 \Bigr]
      - \Bigl(\frac{7}{4} - 7 z + 6 z^2 \Bigr) \ln^2\frac{1 - z}{z}
      + \frac{7+4z}{8}\, \ln^2z
      \nn \\ & \qquad
      + \frac{1}{2} \Li_2(z)
      + \Bigl(\frac{13}{2} - 23 z + \frac{39 z^2}{2} \Bigr) \ln\frac{1 - z}{z}
      + \Bigl(\frac{3}{2} + \frac{9 z}{4} \Bigr) \ln z
      - \frac{34 - 145 z + 121 z^2}{4}
      \nn \\ & \qquad
      + (3 - 12 z + 10 z^2) \frac{\pi^2}{6}
   \biggr\}
   \nn \\ & \quad
   + C_A \biggl\{
      - 2 (1 + 4 z)\, T_3(z)
      - P_{qg}(z) \Bigl[U_3(z) - \frac{1}{6} \ln^3(1-z) + \frac{\pi^2}{6} \ln(1-z) - \frac{\pi^2}{3} \ln z
      \nn \\ & \qquad
         + \frac{7\zeta_3}{2} \Bigr]
      + P_{qg}(-z)\,  S_3(z)
      - z (1 + z)\, S_2(z)
      - 2 z (1 - z) \ln(1-z) \ln z
      \nn \\ & \qquad
      + \Bigl(\frac{2}{3 z} + \frac{1}{2} + 3 z - \frac{25 z^2}{6} \Bigr) \ln^2\frac{1 - z}{z}
      - \Bigl(\frac{15}{4} + \frac{2}{3 z} + 2 z + \frac{47 z^2}{3} \Bigr) \ln^2z
      \nn \\ & \qquad
      - \Bigl(\frac{4}{3 z} + 3 + 8 z + \frac{44 z^2}{3} \Bigr) \Li_2(z)
      + \Bigl(\frac{26}{9 z} + 4 + \frac{31 z}{2} + \frac{50 z^2}{9} \Bigr) \ln(1-z)
      \nn \\ & \qquad
      - \Bigl(\frac{49}{6} + \frac{4 z}{3} + \frac{323 z^2}{18} \Bigr) \ln\frac{1 - z}{z}
      + \frac{104}{27 z} - \frac{19}{36} + \frac{40 z}{9} - \frac{947 z^2}{108}
      \nn \\ & \qquad
      + (2 + z + 24 z^2) \frac{\pi^2}{6}
   \biggr\}
\,.\end{align}
%%%
For simplicity we have suppressed the overall $\theta(1-z)$ multiplying the regular terms. To write the above results in a compact form, we have defined the auxiliary functions,
%%%
\begin{align} \label{eq:Sdef}
S_2(z) &= -2 \Li_2(-z) - 2 \ln(1+z)\ln z - \frac{\pi^2}{6}
\,, \nn \\
S_3(z) 
&= 2 \Li_3(1 - z) - \Li_3(z) + 4 \Li_3\Bigl(\frac{1}{1 + z}\Bigr) - \Li_3(1 - z^2)
   + \frac{\pi^2}{3} \ln(1 + z) - \frac{2}{3} \ln^3(1 + z)
   \nn \\ & \quad
   - \frac{5\zeta_3}{2} + \frac{\pi^2}{6} \ln z + S_2(z) \ln\frac{1-z}{z} - \frac{\ln^3z}{4}
\,, \nn \\
T_3(z)
&=  \Li_3(1 - z) - \Li_2(1 - z)\,\ln(1 - z)
   - \Bigl[\Li_2(z) + \frac{1}{2} \ln^2(1 - z) + \frac{5}{12} \ln^2z - \frac{\pi^2}{3} \Bigr] \ln z
\,, \nn \\
U_3(z)
&= -4 \Li_3(1 - z) + \Li_3(z) - \zeta_3 - \ln(1 - z)\Bigl[\Li_2(z) - \frac{\pi^2}{6} \Bigr] + 2 \Li_2(1 - z)\, \ln z - \frac{\ln^3z}{4}
\,, \nn \\
V_3(z)
&= -4 \Li_3(1 - z) - 5 \Li_3(z) + 5 \zeta_3 + \frac{1}{2} \ln(1 - z) \ln^2z
   \nn \\ & \quad
   - \Bigl[2 \ln^2(1 - z) + \frac{11}{12} \ln^2z -\frac{13\pi^2}{6} \Bigr] \ln z
\,,\end{align}
%%%
which all vanish for $z\to 1$ at least like $1-z$.

Following the discussion in \subsec{matchdet} we can extract the two-loop quark splitting functions $P^{(1)}_{qi}$ from our calculation, and we find agreement with the results of \mycites{Curci:1980uw, Furmanski:1980cm, Ellis:1996nn} using both axial and Feynman gauge.

Our results are also the last important ingredient to obtain the quark beam function at N$^3$LL order, which accounts for all N$^3$LL collinear ISR effects. The all-order structure of the beam function RGE has been discussed in \mycite{Stewart:2010qs}. In addition to our two-loop matching results, the other required ingredients at this order are the three-loop noncusp anomalous dimension, which is known from the analysis in \mycite{Stewart:2010qs}, and the four-loop cusp anomalous dimension. The four-loop correction to the cusp anomalous dimension is not yet known, but can be expected to have an almost negligible numerical impact (similar to what has been observed in thrust~\cite{Becher:2008cf, Abbate:2010xh}).

%%%%%%%%%%%%%%%%%%%%%%%%%%%%%%%%%%%%%%%%%%%%%%%%%%%%%%%%%%%%%%%%%%%%%%%%%%%%%%%%
\section{Conclusions} 
\label{sec:conclusions}
%%%%%%%%%%%%%%%%%%%%%%%%%%%%%%%%%%%%%%%%%%%%%%%%%%%%%%%%%%%%%%%%%%%%%%%%%%%%%%%%

We have calculated at two-loop order the perturbative matching coefficients $\cI_{qj}(t,z,\mu)$
between the virtuality-dependent quark beam function $B_{q}(t, z, \mu)$ and the PDFs $f_{j}(z,\mu)$.
We have performed the calculation using two different
methods and in two different gauges -- covariant Feynman and axial light-cone gauge -- with both methods and gauges yielding 
the same result. The two methods differ in their procedure for taking the discontinuities of the 
operator diagrams that are required to obtain the partonic beam function matrix elements: in the first method the
discontinuity is taken immediately using the Cutkosky rules following \mycites{Curci:1980uw, Ellis:1996nn} whilst in the second the discontinuity is taken after most of the loop integrals have been performed, following \mycite{Stewart:2010qs}. The calculational effort to determine the NNLO matching coefficients is similar for both methods.

Our calculation provides an explicit verification at two loops of the all-orders result~\cite{Stewart:2010qs} that the beam and jet function anomalous dimensions are equal. Conversely, relying on this fact, we are able to extract the two-loop quark splitting functions, $P_{qi}$, and find agreement with the well-known results~\cite{Ellis:1996nn, Furmanski:1980cm}.
Our results are an important ingredient to obtain the NNLO singular contributions as well as the N$^3$LL resummation for observables that probe the virtuality of the colliding partons, such as $N$-jettiness.

%%%%%%%%%%%%%%%%%%%%%%%%%%%%%%%%%%%%%%%%%%%%%%%%%%%%%%%%%%%%%%%%%%%%%%%%%%%%%%%%
\begin{acknowledgments}
We like to thank Wouter Waalewijn, Jonathan Walsh, and Iain Stewart for helpful discussions.
The Feynman diagrams in this paper have been drawn using {\tt JaxoDraw}~\cite{Binosi:2008ig}.
This work was supported by the DFG Emmy-Noether Grant No. TA 867/1-1.
\end{acknowledgments}
%%%%%%%%%%%%%%%%%%%%%%%%%%%%%%%%%%%%%%%%%%%%%%%%%%%%%%%%%%%%%%%%%%%%%%%%%%%%%%%%

\appendix

%%%%%%%%%%%%%%%%%%%%%%%%%%%%%%%%%%%%%%%%%%%%%%%%%%%%%%%%%%%%%%%%%%%%%%%%%%%%%%%%
\section{Anomalous dimensions and two-loop matching ingredients}
\label{app:Formulae}
%%%%%%%%%%%%%%%%%%%%%%%%%%%%%%%%%%%%%%%%%%%%%%%%%%%%%%%%%%%%%%%%%%%%%%%%%%%%%%%%

\subsection{Anomalous dimensions}

We define the expansion of the beam function noncusp anomalous dimension as
\begin{equation}
\gamma^i_B(\alpha_s) = \sum_{n=0}^\infty \gamma_{B\,n}^i \Bigl(\frac{\alpha_s}{4\pi}\Bigr)^{n+1}
\,.\end{equation}
For the quark beam function in $\overline{\mathrm{MS}}$ we have~\cite{Stewart:2010qs}
%%%
\begin{align} \label{eq:gammaBqexp}
\gamma_{B\,0}^q &= 6 C_F
\,, \nn \\
\gamma_{B\,1}^q
&= C_F \Bigl[
   C_A \Bigl(\frac{146}{9} - 80 \zeta_3\Bigr)
   + C_F (3 - 4 \pi^2 + 48 \zeta_3)
   + \beta_0 \Bigl(\frac{121}{9} + \frac{2\pi^2}{3} \Bigr) \Bigr]
\,, \nn \\
\gamma_{B\,2}^q
&= 2 C_F \Bigl[
   C_A^2 \Bigl(\frac{52019}{162} - \frac{841\pi^2}{81} - \frac{82\pi^4}{27} -\frac{2056\zeta_3}{9}
      + \frac{88\pi^2 \zeta_3}{9} + 232 \zeta_5\Bigr)
   \nn\\ & \qquad
   + C_A C_F \Bigl(\frac{151}{4} - \frac{205\pi^2}{9} - \frac{247\pi^4}{135} + \frac{844\zeta_3}{3}
      + \frac{8\pi^2 \zeta_3}{3} + 120 \zeta_5\Bigr)
   \nn\\ & \qquad
   + C_F^2 \Bigl(\frac{29}{2} + 3 \pi^2 + \frac{8\pi^4}{5} + 68 \zeta_3 - \frac{16\pi^2 \zeta_3}{3} - 240 \zeta_5\Bigr)
   \nn\\ & \qquad
   +  C_A\beta_0 \Bigl(-\frac{7739}{54} + \frac{325}{81} \pi^2 + \frac{617 \pi^4}{270} - \frac{1276\zeta_3}{9} \Bigr)
   \nn\\ & \qquad
   + \beta_0^2 \Bigl(-\frac{3457}{324} + \frac{5\pi^2}{9} + \frac{16 \zeta_3}{3} \Bigr)
   + \beta_1 \Bigl(\frac{1166}{27} - \frac{8 \pi^2}{9} - \frac{41 \pi^4}{135} + \frac{52 \zeta_3}{9}\Bigr)
   \Bigr]
\,.\end{align}
%%%
We write the expansion of the cusp anomalous dimension as
\begin{align}
\Gamma^i_\cusp(\alpha_s) = \sum_{n=0}^\infty \Gamma^i_n \Bigl(\frac{\alpha_s}{4\pi}\Bigr)^{n+1}
\,.\end{align}
The coefficients of the $\overline{\mathrm{MS}}$ cusp anomalous dimension to three loops are~\cite{Korchemsky:1987wg, Moch:2004pa}
%%%
\begin{align} \label{eq:Gacuspexp}
\Gamma^q_i &= C_F \Gamma_i
\,,\qquad
\Gamma^g_i = C_A \Gamma_i
\,,\qquad \text{(for $i = 0,1,2$)}
\,,\nn\\[1ex]
\Gamma_0 &= 4
\,,\nn\\
\Gamma_1
&= 4 \Bigl[ C_A \Bigl( \frac{67}{9} - \frac{\pi^2}{3} \Bigr)  - \frac{20}{9}\,T_F\, n_f \Bigr]
= \frac{4}{3} \bigl[ (4 - \pi^2) C_A + 5 \beta_0 \bigr]
\,,\nn\\
\Gamma_2
&= 4 \Bigl[
   C_A^2 \Bigl(\frac{245}{6} -\frac{134 \pi^2}{27} + \frac{11 \pi ^4}{45} + \frac{22 \zeta_3}{3}\Bigr)
   +  C_A\, T_F\,n_f \Bigl(- \frac{418}{27} + \frac{40 \pi^2}{27}  - \frac{56 \zeta_3}{3} \Bigr)
   \nn \\ & \qquad
   +  C_F\, T_F\,n_f \Bigl(- \frac{55}{3} + 16 \zeta_3 \Bigr)
   - \frac{16}{27}\,T_F^2\, n_f^2
   \Bigr]
\,.\end{align}
%%%

%===============================================================================
\subsection{One-loop beam function matching coefficients}
%===============================================================================

We define the expansion of the beam function matching coefficient as follows:
\begin{align}
\cI_{ji} = \sum_{n=0}^{\infty} \biggl(\dfrac{\alpha_s}{4\pi}\biggr)^n \cI_{ji}^{(n)}
\,.\end{align}
The tree-level matching coefficients are
%%%
\begin{equation}
\cI_{ij}^\zero(t,z,\mu) = \delta(t)\, \delta_{ij}\delta(1-z)
\,.\end{equation}
%%%
The one-loop matching coefficients are
%%%
\begin{align} \label{eq:Iq}
\cI_{ij}^\one(t,z,\mu)
&= \frac{1}{\mu^2} \cL_1\Bigl(\frac{t}{\mu^2}\Bigr) \Gamma_0^i\, \delta_{ij}\delta(1 - z)
  + \frac{1}{\mu^2} \cL_0\Bigl(\frac{t}{\mu^2}\Bigr)
  \Bigl[- \frac{\gamma_{B\,0}^i}{2}\,\delta_{ij}\delta(1-z) + 2 P_{ij}^\zero(z) \Bigr]
  \nn \\ & \quad
  + \delta(t)\, 2I_{ij}^\one(z)
\,.\end{align}
%%%
%
The $\mu$-independent one-loop constants are
%%%
\begin{align}
I_{q_iq_j}^\one(z) &= \delta_{ij}\, C_F\, \theta(z) I_{qq}(z)
\,,\nn\\
I_{q_ig}^\one(z) &= T_F\, \theta(z) I_{qg}(z)
\,,\nn\\
I_{gg}^\one(z) &= C_A\, \theta(z) I_{gg}(z)
\,,\nn\\
I_{gq_i}^\one(z) &= C_F\, \theta(z) I_{gq}(z)
\,,\end{align}
%%%
with the quark matching functions~\cite{Stewart:2010qs} given by\footnote{Note that here $I_{ij}(z) \equiv \cI_{ij}^{(1,\delta)}(z)$ in the notation of \mycites{Stewart:2010qs, Berger:2010xi}.}
%%%
\begin{align} \label{eq:Iqdel_results}
I_{qq}(z)
&= \cL_1(1 - z)(1 + z^2)
  -\frac{\pi^2}{6}\, \delta(1 - z)
  + \theta(1 - z)\Bigl(1 - z - \frac{1 + z^2}{1 - z}\ln z \Bigr)
\,, \nn \\
I_{qg}(z)
&= P_{qg}(z)\Bigl(\ln\frac{1-z}{z} - 1\Bigr) +  \theta(1-z)
\,.\end{align}
%%%

%===============================================================================
\subsection{Splitting functions}
%===============================================================================

We define the expansion of PDF anomalous dimensions ($\gamma^f_{ij}=2 P_{ij}$) in the $\overline{\mathrm{MS}}$ as follows:
\begin{align}
\label{eq:Pijexp}
P_{ij}(z,\alpha_s) = \sum_{n=0}^{\infty} \left(\dfrac{\alpha_s}{2\pi}\right)^{n+1} P_{ij}^{(n)}(z)\,.
\end{align}
The one-loop terms read
%%%
\begin{align}
P_{q_i q_j}^\zero(z) &= C_F\, \theta(z)\, \delta_{ij} P_{qq}(z)
\,,\nn\\
P_{q_ig}^\zero(z) = P_{\bar q_ig}^\zero(z) &= T_F\, \theta(z) P_{qg}(z)
\,,\nn\\
P_{gg}^\zero(z) &= C_A\, \theta(z) P_{gg}(z) + \frac{\beta_0}{2}\,\delta(1-z)
\,,\nn\\
P_{gq_i}^\zero(z) = P_{g\bar q_i}^\zero(z) &= C_F\, \theta(z) P_{gq}(z)
\,,\end{align}
%%%
with the usual one-loop (LO) quark and gluon splitting functions
%%%
\begin{align} \label{eq:Pij}
P_{qq}(z)
&= \cL_0(1-z)(1+z^2) + \frac{3}{2}\,\delta(1-z)
\equiv \biggl[\theta(1-z)\,\frac{1+z^2}{1-z}\biggr]_+
\,,\nn\\
P_{qg}(z) &= \theta(1-z)\bigl[(1-z)^2+ z^2\bigr]
\,,\nn\\
P_{gg}(z)
&= 2 \cL_0(1-z) \frac{(1 - z + z^2)^2}{z}
\,,\nn\\
P_{gq}(z) &= \theta(1-z)\, \frac{1+(1-z)^2}{z}
\,.\end{align}
%%%
%
At two loops we have
%%%
\begin{align}
P_{q_i q_j}^\one(z) &= C_F\, \theta(z) \bigl[ \delta_{ij} P_{qqV}^\one(z) + P_{qqS}^\one(z) \bigr]
\,, \nn \\
P_{q_i \bar q_j}^\one(z) &= C_F\, \theta(z) \bigl[ \delta_{ij} P_{q\bar qV}^\one(z) + P_{qqS}^\one(z) \bigr]
\,, \nn \\
P_{q_i g}^\one(z) = P_{\bar q_i g}^\one(z) &= T_F\, \theta(z) P_{qg}^\one(z)
\,.\end{align}
%%%
The two-loop (NLO) splitting functions were calculated in \mycite{Furmanski:1980cm, Ellis:1996nn}. Using the results of \mycite{Ellis:1996nn}, we have for the splitting functions we need for the matching:
%%%
\begin{align}
P_{qqV}^\one(z)
&= \frac{\Gamma_1}{8} \cL_0(1-z) (1 + z^2)
   \nn \\ & \quad
   + \delta(1-z) \biggl[ C_F \Bigl(\frac{3}{8} - \frac{\pi^2}{2} + 6\zeta_3\Bigr) + C_A \Bigl(\frac{1}{4} - 3 \zeta_3\Bigr)
      + \beta_0 \Bigl(\frac{1}{8} + \frac{\pi^2}{6}\Bigr) \biggr]
   \nn \\ & \quad
   - C_F \biggl\{\frac{1 + z^2}{1 - z} \Bigl[2\ln(1-z) + \frac{3}{2}\Bigr]\ln z + \frac{1 + z}{2} \ln^2 z  + \frac{3 + 7 z}{2}\ln z + 5 (1 - z) \biggr\}
   \nn \\ & \quad
   + C_A \biggl[\frac{1}{2}\, \frac{1 + z^2}{1 - z} \ln^2 z + (1 + z) \ln z + 3 (1 - z) \biggr]
%    \nn \\ & \quad
   + \beta_0 \biggl[\frac{1}{2}\,\frac{1 + z^2}{1 - z}\ln z + 1 - z \biggr]
\,, \nn \\
P_{q\bar qV}^\one(z)
&= (2 C_F - C_A) \biggl\{\frac{1 + z^2}{1 + z} \Bigl[S_2(z) + \frac{1}{2}\ln^2 z \Bigr] + (1 + z) \ln z + 2 (1 - z) \biggr\}
% \nn \\
% &= (2 C_F - C_A) \biggl\{\frac{1 + z^2}{1 + z} \Bigl[-2 \Li_2(-z) - 2 \ln(1+z)\ln z - \frac{\pi^2}{6} + \frac{1}{2}\ln^2 z \Bigr]
%    + (1 + z) \ln z + 2 (1 - z) \biggr\}
\,, \nn \\
P_{qqS}^\one(z)
&= T_F \biggl[- (1 + z) \ln^2 z + \Bigl(1 + 5 z + \frac{8}{3} z^2 \Bigr) \ln z + \frac{20}{9z} - 2 + 6 z - \frac{56}{9} z^2 \biggr]
\,,\end{align}
%%%
and
%%%
\begin{align} \label{eq:Pqg_one}
P_{qg}^\one(z)
&= C_F \biggl[
      P_{qg}(z) \Bigl(\ln^2\frac{1-z}{z} - 2 \ln\frac{1-z}{z} - \frac{\pi^2}{3} + 5\Bigr)
      + 2 \ln(1-z)
      \nn \\ & \qquad
      - \frac{1 - 2 z}{2} \ln^2 z - \frac{1 - 4 z}{2} \ln z
      + 2 - \frac{9}{2} z
   \biggr]
   \nn \\ & \quad
   + C_A \biggl\{
      P_{qg}(z) \Bigl[- \ln^2(1-z) + 2 \ln(1-z) +\frac{22}{3}\ln z - \frac{109}{9} + \frac{\pi^2}{6} \Bigr]
      + P_{qg}(-z) S_2(z)
      \nn \\ & \qquad
      - 2 \ln(1-z) - (1 + 2 z) \ln^2 z  + \frac{68 z - 19}{3}\ln z
      + \frac{20}{9 z} + \frac{91}{9} + \frac{7}{9} z
   \biggr\}
\,.\end{align}
%%%
The auxiliary function $S_2(z)$ was already defined in \eq{Sdef}. For simplicity we have suppressed the overall $\theta(1-z)$ multiplying the regular contributions.

\subsection{Convolutions of one-loop functions}

The convolution of two one-loop functions is defined as (the index $j$ is not summed over)
%%%
\begin{equation}
(P_{ij} \conv P_{jk})(z) \equiv P_{ij}(z) \convz P_{jk}(z) = \int_z^1\! \frac{\df w}{w}\, P_{ij}(w) P_{jk} \Bigl(\frac{z}{w}\Bigr)
\,,\end{equation}
%%%
and analogously for $(I_{ij} \conv P_{jk})(z)$.

The convolutions of the splitting functions in \eq{Pij} that we require are:
%%%
\begin{align}
(P_{qq} \conv P_{qq})(z)
&= 4\cL_1(1-z)(1+z^2)
  + 3P_{qq}(z) - \Bigl(\frac{9}{4} + \frac{2\pi^2}{3}\Bigr) \delta(1-z)
  \nn \\ & \quad
  + \bigl[- 2 P_{qq}(z) + 1+z\bigr] \ln z
  - 2(1 - z)
\,,\nn\\
(P_{qg} \conv P_{gq})(z)
&= 2(1+z)\ln z
  + \frac{4}{3z} + 1 - z - \frac{4}{3}z^2
\,,\nn\\
(P_{qg} \conv P_{gg})(z)
% &= 1+\frac{4}{3z}+8z-\frac{31z^2}{3}+2(1-2z+2z^2)\ln(1-z)+2(1+4z)\ln z
% \nn\\
&= 2 P_{qg}(z) \ln(1-z)
  + 2(1 + 4z) \ln z
  + \frac{4}{3z} + 1 + 8z - \frac{31}{3} z^2
\,,\nn\\
(P_{qq}\conv P_{qg})(z)
% &= -\frac{1}{2}+2z+2(1-2z+2z^2)\ln(1-z)+(-1+2z-4z^2)\ln z
% \nn\\
&= 2 P_{qg}(z)\ln\frac{1-z}{z}
  + (1 - 2z) \ln z
  - \frac{1}{2} + 2z
\,.\end{align}
%%%

The convolutions that we require of the one-loop matching functions $I_{ij}(z)$ in
\eq{Iqdel_results} with the one-loop splitting functions $P_{jk}(z)$ in \eq{Pij}  are:
%%%
\begin{align}
(I_{qq} \conv P_{qq})(z)
&= \Bigl\{3 \cL_2(1-z)
  + \cL_0(1-z) \Bigl[\ln^2 z - 4 \ln(1-z) \ln z - \frac{\pi^2}{2} \Bigr] \Bigr\}(1+z^2)
  + \frac{3}{2} I_{qq}(z)
  \nn \\ & \quad
  + 4\zeta_3 \delta(1-z)
  - (1 + z) \Bigl[\Li_2(z) + \frac{1}{2} \ln^2 z - \frac{\pi^2}{6} \Bigr]
  - z \ln z - 1 + z
\,, \nn \\
(I_{qg} \conv P_{gq})(z)
&= -2(1 + z) \Bigl[\Li_2(z) + \frac{1}{2} \ln^2 z - \frac{\pi^2}{6} \Bigr]
  - \Bigl(1 + z + \frac{4}{3} z^2 \Bigr) \ln\frac{1-z}{z}
  \nn \\ & \quad
  + \Bigl(\frac{4}{3z} + 2 \Bigr) \ln(1-z)
  + \frac{1}{3}\Bigl(\frac{2}{z} - 5 - z + 4 z^2\Bigr)
\,, \nn \\
(I_{qg} \conv P_{gg})(z)
&= 2 P_{qg}(z) \Bigr[\ln(1-z) \ln\frac{1-z}{z}-\frac{\pi^2}{6} \Bigr]
  - 2(1 + 4z) \Bigl[\Li_2(z) + \frac{1}{2} \ln^2 z - \frac{\pi^2}{6} \Bigr]
  \nn \\ & \quad
  + \Bigl(\frac{4}{3z} + 1 + 12 z - \frac{43}{3} z^2 \Bigr) \ln(1-z)
  + \Bigl(1 - 8z + \frac{31}{3} z^2\Bigr) \ln z
  \nn \\ & \quad
  +  \frac{2}{3z} - \frac{13}{6} - \frac{37}{3} z + \frac{83}{6} z^2
\,, \nn \\
(I_{qq} \conv P_{qg})(z)
&= P_{qg}(z) \Bigl(\ln^2\frac{1-z}{z} - \frac{\pi^2}{6}\Bigr)
  - (1 - 2z) \Bigl[\Li_2(z) + \frac{1}{2} \ln^2 z - \frac{\pi^2}{6} \Bigr]
  \nn \\ & \quad
  + z (7 - 3 z) \ln\frac{1-z}{z}
  - 2(1+z) \ln(1-z)
  -\frac{1}{2} - 4 z + \frac{9}{2}z^2
\,.\end{align}

%%%%%%%%%%%%%%%%%%%%%%%%%%%%%%%%%%%%%%%%%%%%%%%%%%%%%%%%%%%%%%%%%%%%%%%%%%%%%%%%
\section{Plus distributions and discontinuities}
\label{app:Useful}
%%%%%%%%%%%%%%%%%%%%%%%%%%%%%%%%%%%%%%%%%%%%%%%%%%%%%%%%%%%%%%%%%%%%%%%%%%%%%%%%

We define the standard plus distributions as
\begin{align} \label{eq:plusdef}
\cL_n(x)
&= \biggl[ \frac{\theta(x) \ln^n x}{x}\biggr]_+
 = \lim_{\eps \to 0} \frac{\df}{\df x}\biggl[ \theta(x- \eps)\frac{\ln^{n+1} x}{n+1} \biggr]
\,.\end{align}
%%%%
We need the distributional identity
\begin{align} \label{eq:plus_exp}
\frac{\theta(x)}{x^{1-\eps}} = \frac{1}{\eps}\, \delta(x) + \sum_{n = 0}^\infty \frac{\eps^n}{n!}\, \cL_n(x)
= \frac{1}{\eps}\, \delta(x) + \cL_0(x) + \eps \cL_1(x) + \ord{\eps^2}
\,,\end{align}
and the derivatives
%%%
\begin{align}
\mu\frac{\df}{\df\mu}\,\frac{1}{\mu^2}\cL_n\Bigl(\frac{t}{\mu^2}\Bigr)
&= -2n\, \frac{1}{\mu^2}\cL_{n-1}\Bigl(\frac{t}{\mu^2}\Bigr)
\qquad (\forall\, n \ge 1)
\,, \nn \\
\mu\frac{\df}{\df\mu}\,\frac{1}{\mu^2}\cL_0\Bigl(\frac{t}{\mu^2}\Bigr)
&= \delta(t)
\,.\end{align}
%%%
The discontinuity of a function $g(x)$ is defined as
%%%
\begin{align} \label{eq:Disc_def2}
\Disc_x\, g(x) = \lim_{\beta\to 0} \bigl[ g(x + \img\beta) - g(x - \img\beta) \bigr]
\,.\end{align}
%
%If $g(x)$ is real then $\Disc_x g(x) = 2\img\, \Im\, g(x+\img0)$. 
In the Dispersive Method we determine the discontinuity (with respect to $t$) of a two-loop Feynman graph employing the identity (see e.g. \mycite{Stewart:2010qs})
%%%
\begin{align} \label{eq:disc_os2}
\frac{\img}{2\pi} \Disc_x\, (-x)^{n-\eps}
= (-1)^{n-1} \frac{\sin(\pi\eps)}{\pi}\, \theta(x)\, x^{n-\eps}
\,.\end{align}
%%%

%% Bibliography
\bibliographystyle{../jhep}
\bibliography{../beamfunc}

\providecommand{\href}[2]{#2}\begingroup\raggedright\begin{thebibliography}{10}

\bibitem{Stewart:2010qs}
I.~W. Stewart, F.~J. Tackmann, and W.~J. Waalewijn, {\it {The Quark Beam
  Function at NNLL}},  {\em JHEP} {\bf 1009} (2010) 005,
  [\href{http://arXiv.org/abs/1002.2213}{{\tt arXiv:1002.2213}}].
%%CITATION = ARXIV:1002.2213;%%

\bibitem{Collins:1988ig}
J.~C. Collins, D.~E. Soper, and G.~F. Sterman, {\it {Soft Gluons and
  Factorization}},  {\em Nucl.~Phys.} {\bf B308} (1988) 833.
%%CITATION = NUPHA,B308,833;%%

\bibitem{Bodwin:1984hc}
G.~T. Bodwin, {\it {Factorization of the Drell-Yan Cross-Section in
  Perturbation Theory}},  {\em Phys.~Rev.~} {\bf 31} (1985) 2616.
%%CITATION = PHRVA,D31,2616;%%

\bibitem{Collins:1985ue}
J.~C. Collins, D.~E. Soper, and G.~F. Sterman, {\it {Factorization for Short
  Distance Hadron - Hadron Scattering}},  {\em Nucl.~Phys.} {\bf B261} (1985)
  104.
%%CITATION = NUPHA,B261,104;%%

\bibitem{Stewart:2009yx}
I.~W. Stewart, F.~J. Tackmann, and W.~J. Waalewijn, {\it {Factorization at the
  LHC: From PDFs to Initial State Jets}},  {\em Phys.~Rev.~D} {\bf 81} (2010)
  094035, [\href{http://arXiv.org/abs/0910.0467}{{\tt arXiv:0910.0467}}].
%%CITATION = ARXIV:0910.0467;%%

\bibitem{Collins:1984kg}
J.~C. Collins, D.~E. Soper, and G.~F. Sterman, {\it {Transverse Momentum
  Distribution in Drell-Yan Pair and W and Z Boson Production}},  {\em
  Nucl.~Phys.} {\bf B250} (1985) 199.
%%CITATION = NUPHA,B250,199;%%

\bibitem{Collins:1350496}
J.~C. Collins, {\em Foundations of perturbative QCD}.
\newblock Cambridge monographs on particle physics, nuclear physics, and
  cosmology. Cambridge Univ. Press, New York, NY, 2011.

\bibitem{Chiu:2012ir}
J.-Y. Chiu, A.~Jain, D.~Neill, and I.~Z. Rothstein, {\it {A Formalism for the
  Systematic Treatment of Rapidity Logarithms in Quantum Field Theory}},  {\em
  JHEP} {\bf 1205} (2012) 084, [\href{http://arXiv.org/abs/1202.0814}{{\tt
  arXiv:1202.0814}}].
%%CITATION = ARXIV:1202.0814;%%

\bibitem{Becher:2010tm}
T.~Becher and M.~Neubert, {\it {Drell-Yan production at small $q_T$, transverse
  parton distributions and the collinear anomaly}},  {\em Eur.~Phys.~J.~C} {\bf
  71} (2011) 1665, [\href{http://arXiv.org/abs/1007.4005}{{\tt
  arXiv:1007.4005}}].
%%CITATION = ARXIV:1007.4005;%%

\bibitem{Catani:2000vq}
S.~Catani, D.~de~Florian, and M.~Grazzini, {\it {Universality of nonleading
  logarithmic contributions in transverse momentum distributions}},  {\em
  Nucl.~Phys.} {\bf B596} (2001) 299--312,
  [\href{http://arXiv.org/abs/hep-ph/0008184}{{\tt hep-ph/0008184}}].
%%CITATION = HEP-PH/0008184;%%

\bibitem{Catani:2010pd}
S.~Catani and M.~Grazzini, {\it {QCD transverse-momentum resummation in gluon
  fusion processes}},  {\em Nucl.~Phys.} {\bf B845} (2011) 297--323,
  [\href{http://arXiv.org/abs/1011.3918}{{\tt arXiv:1011.3918}}].
%%CITATION = ARXIV:1011.3918;%%

\bibitem{GarciaEchevarria:2011rb}
M.~G. Echevarria, A.~Idilbi, and I.~Scimemi, {\it {Factorization Theorem For
  Drell-Yan At Low $q_T$ And Transverse Momentum Distributions
  On-The-Light-Cone}},  {\em JHEP} {\bf 1207} (2012) 002,
  [\href{http://arXiv.org/abs/1111.4996}{{\tt arXiv:1111.4996}}].
%%CITATION = ARXIV:1111.4996;%%

\bibitem{Echevarria:2012js}
M.~G. Echevarría, A.~Idilbi, and I.~Scimemi, {\it {Soft and Collinear
  Factorization and Transverse Momentum Dependent Parton Distribution
  Functions}},  {\em Phys.~Lett.~B} {\bf 726} (2013) 795--801,
  [\href{http://arXiv.org/abs/1211.1947}{{\tt arXiv:1211.1947}}].
%%CITATION = ARXIV:1211.1947;%%

\bibitem{Becher:2012yn}
T.~Becher, M.~Neubert, and D.~Wilhelm, {\it {Higgs-Boson Production at Small
  Transverse Momentum}},  {\em JHEP} {\bf 1305} (2013) 110,
  [\href{http://arXiv.org/abs/1212.2621}{{\tt arXiv:1212.2621}}].
%%CITATION = ARXIV:1212.2621;%%

\bibitem{Fleming:2006cd}
S.~Fleming, A.~K. Leibovich, and T.~Mehen, {\it {Resummation of Large Endpoint
  Corrections to Color-Octet $J/\psi$ Photoproduction}},  {\em Phys.~Rev.~D}
  {\bf 74} (2006) 114004, [\href{http://arXiv.org/abs/hep-ph/0607121}{{\tt
  hep-ph/0607121}}].
%%CITATION = HEP-PH/0607121;%%

\bibitem{Stewart:2010tn}
I.~W. Stewart, F.~J. Tackmann, and W.~J. Waalewijn, {\it {N-Jettiness: An
  Inclusive Event Shape to Veto Jets}},  {\em Phys.~Rev.~Lett.} {\bf 105}
  (2010) 092002, [\href{http://arXiv.org/abs/1004.2489}{{\tt
  arXiv:1004.2489}}].
%%CITATION = ARXIV:1004.2489;%%

\bibitem{Jouttenus:2011wh}
T.~T. Jouttenus, I.~W. Stewart, F.~J. Tackmann, and W.~J. Waalewijn, {\it {The
  Soft Function for Exclusive N-Jet Production at Hadron Colliders}},  {\em
  Phys.~Rev.~D} {\bf 83} (2011) 114030,
  [\href{http://arXiv.org/abs/1102.4344}{{\tt arXiv:1102.4344}}].
%%CITATION = ARXIV:1102.4344;%%

\bibitem{Stewart:2010pd}
I.~W. Stewart, F.~J. Tackmann, and W.~J. Waalewijn, {\it {The Beam Thrust Cross
  Section for Drell-Yan at NNLL Order}},  {\em Phys.~Rev.~Lett.} {\bf 106}
  (2011) 032001, [\href{http://arXiv.org/abs/1005.4060}{{\tt
  arXiv:1005.4060}}].
%%CITATION = ARXIV:1005.4060;%%

\bibitem{Berger:2010xi}
C.~F. Berger, C.~Marcantonini, I.~W. Stewart, F.~J. Tackmann, and W.~J.
  Waalewijn, {\it {Higgs Production with a Central Jet Veto at NNLL+NNLO}},
  {\em JHEP} {\bf 1104} (2011) 092, [\href{http://arXiv.org/abs/1012.4480}{{\tt
  arXiv:1012.4480}}].
%%CITATION = ARXIV:1012.4480;%%

\bibitem{Jouttenus:2013hs}
T.~T. Jouttenus, I.~W. Stewart, F.~J. Tackmann, and W.~J. Waalewijn, {\it {Jet
  Mass Spectra in Higgs $+$ One Jet at NNLL}},  {\em Phys.~Rev.~D} {\bf 88}
  (2013) 054031, [\href{http://arXiv.org/abs/1302.0846}{{\tt
  arXiv:1302.0846}}].
%%CITATION = ARXIV:1302.0846;%%

\bibitem{Kang:2013nha}
D.~Kang, C.~Lee, and I.~W. Stewart, {\it {Using 1-Jettiness to Measure 2 Jets
  in DIS 3 Ways}},  {\em Phys.~Rev.~D} {\bf 88} (2013) 054004,
  [\href{http://arXiv.org/abs/1303.6952}{{\tt arXiv:1303.6952}}].
%%CITATION = ARXIV:1303.6952;%%

\bibitem{Kang:2013lga}
Z.-B. Kang, X.~Liu, and S.~Mantry, {\it {The 1-Jettiness DIS event shape: NNLL
  + NLO results}},  \href{http://arXiv.org/abs/1312.0301}{{\tt
  arXiv:1312.0301}}.
%%CITATION = ARXIV:1312.0301;%%

\bibitem{Kang:2012zr}
Z.-B. Kang, S.~Mantry, and J.-W. Qiu, {\it {N-Jettiness as a Probe of Nuclear
  Dynamics}},  {\em Phys.~Rev.~D} {\bf 86} (2012) 114011,
  [\href{http://arXiv.org/abs/1204.5469}{{\tt arXiv:1204.5469}}].
%%CITATION = ARXIV:1204.5469;%%

\bibitem{Kang:2013wca}
Z.-B. Kang, X.~Liu, S.~Mantry, and J.-W. Qiu, {\it {Probing nuclear dynamics in
  jet production with a global event shape}},  {\em Phys.~Rev.~D} {\bf 88}
  (2013) 074020, [\href{http://arXiv.org/abs/1303.3063}{{\tt
  arXiv:1303.3063}}].
%%CITATION = ARXIV:1303.3063;%%

\bibitem{Collins:2007ph}
J.~Collins, T.~Rogers, and A.~Stasto, {\it {Fully unintegrated parton
  correlation functions and factorization in lowest-order hard scattering}},
  {\em Phys.~Rev.~D} {\bf 77} (2008) 085009,
  [\href{http://arXiv.org/abs/0708.2833}{{\tt arXiv:0708.2833}}].
%%CITATION = ARXIV:0708.2833;%%

\bibitem{Rogers:2008jk}
T.~C. Rogers, {\it {Next-to-Leading Order Hard Scattering Using Fully
  Unintegrated Parton Distribution Functions}},  {\em Phys.~Rev.~D} {\bf 78}
  (2008) 074018, [\href{http://arXiv.org/abs/0807.2430}{{\tt
  arXiv:0807.2430}}].
%%CITATION = ARXIV:0807.2430;%%

\bibitem{Mantry:2009qz}
S.~Mantry and F.~Petriello, {\it {Factorization and Resummation of Higgs Boson
  Differential Distributions in Soft-Collinear Effective Theory}},  {\em
  Phys.~Rev.~D} {\bf 81} (2010) 093007,
  [\href{http://arXiv.org/abs/0911.4135}{{\tt arXiv:0911.4135}}].
%%CITATION = ARXIV:0911.4135;%%

\bibitem{Mantry:2010mk}
S.~Mantry and F.~Petriello, {\it {Transverse Momentum Distributions from
  Effective Field Theory with Numerical Results}},  {\em Phys.~Rev.~D} {\bf 83}
  (2011) 053007, [\href{http://arXiv.org/abs/1007.3773}{{\tt
  arXiv:1007.3773}}].
%%CITATION = ARXIV:1007.3773;%%

\bibitem{Jain:2011iu}
A.~Jain, M.~Procura, and W.~J. Waalewijn, {\it {Fully-Unintegrated Parton
  Distribution and Fragmentation Functions at Perturbative $k_T$}},  {\em JHEP}
  {\bf 1204} (2012) 132, [\href{http://arXiv.org/abs/1110.0839}{{\tt
  arXiv:1110.0839}}].
%%CITATION = ARXIV:1110.0839;%%

\bibitem{Mantry:2010bi}
S.~Mantry and F.~Petriello, {\it {Transverse Momentum Distributions in the
  Non-Perturbative Region}},  {\em Phys.~Rev.~D} {\bf 84} (2011) 014030,
  [\href{http://arXiv.org/abs/1011.0757}{{\tt arXiv:1011.0757}}].
%%CITATION = ARXIV:1011.0757;%%

\bibitem{Becher:2012qa}
T.~Becher and M.~Neubert, {\it {Factorization and NNLL Resummation for Higgs
  Production with a Jet Veto}},  {\em JHEP} {\bf 1207} (2012) 108,
  [\href{http://arXiv.org/abs/1205.3806}{{\tt arXiv:1205.3806}}].
%%CITATION = ARXIV:1205.3806;%%

\bibitem{Tackmann:2012bt}
F.~J. Tackmann, J.~R. Walsh, and S.~Zuberi, {\it {Resummation Properties of Jet
  Vetoes at the LHC}},  {\em Phys.~Rev.~D} {\bf 86} (2012) 053011,
  [\href{http://arXiv.org/abs/1206.4312}{{\tt arXiv:1206.4312}}].
%%CITATION = ARXIV:1206.4312;%%

\bibitem{Liu:2012sz}
X.~Liu and F.~Petriello, {\it {Resummation of jet-veto logarithms in hadronic
  processes containing jets}},  {\em Phys.~Rev.~D} {\bf 87} (2013) 014018,
  [\href{http://arXiv.org/abs/1210.1906}{{\tt arXiv:1210.1906}}].
%%CITATION = ARXIV:1210.1906;%%

\bibitem{Becher:2013xia}
T.~Becher, M.~Neubert, and L.~Rothen, {\it {Factorization and
  $N^{3}LL_{p}$+NNLO predictions for the Higgs cross section with a jet veto}},
   {\em JHEP} {\bf 1310} (2013) 125,
  [\href{http://arXiv.org/abs/1307.0025}{{\tt arXiv:1307.0025}}].
%%CITATION = ARXIV:1307.0025;%%

\bibitem{Stewart:2013faa}
I.~W. Stewart, F.~J. Tackmann, J.~R. Walsh, and S.~Zuberi, {\it {Jet $p_T$
  Resummation in Higgs Production at NNLL$'$+NNLO}},  {\em Phys.~Rev.~D} {\bf
  89} (2014) 054001, [\href{http://arXiv.org/abs/1307.1808}{{\tt
  arXiv:1307.1808}}].
%%CITATION = ARXIV:1307.1808;%%

\bibitem{Li:2014ria}
Y.~Li and X.~Liu, {\it {High precision predictions for exclusive $VH$
  production at the LHC}},  \href{http://arXiv.org/abs/1401.2149}{{\tt
  arXiv:1401.2149}}.
%%CITATION = ARXIV:1401.2149;%%

\bibitem{Bauer:2000ew}
C.~W. Bauer, S.~Fleming, and M.~E. Luke, {\it {Summing Sudakov logarithms in
  $B\to X_s\gamma$ in effective field theory}},  {\em Phys.~Rev.~D} {\bf 63}
  (2000) 014006, [\href{http://arXiv.org/abs/hep-ph/0005275}{{\tt
  hep-ph/0005275}}].
%%CITATION = HEP-PH/0005275;%%

\bibitem{Bauer:2000yr}
C.~W. Bauer, S.~Fleming, D.~Pirjol, and I.~W. Stewart, {\it {An Effective field
  theory for collinear and soft gluons: Heavy to light decays}},  {\em
  Phys.~Rev.~D} {\bf 63} (2001) 114020,
  [\href{http://arXiv.org/abs/hep-ph/0011336}{{\tt hep-ph/0011336}}].
%%CITATION = HEP-PH/0011336;%%

\bibitem{Bauer:2001ct}
C.~W. Bauer and I.~W. Stewart, {\it {Invariant operators in collinear effective
  theory}},  {\em Phys.~Lett.~B} {\bf 516} (2001) 134--142,
  [\href{http://arXiv.org/abs/hep-ph/0107001}{{\tt hep-ph/0107001}}].
%%CITATION = HEP-PH/0107001;%%

\bibitem{Bauer:2001yt}
C.~W. Bauer, D.~Pirjol, and I.~W. Stewart, {\it {Soft collinear factorization
  in effective field theory}},  {\em Phys.~Rev.~D} {\bf 65} (2002) 054022,
  [\href{http://arXiv.org/abs/hep-ph/0109045}{{\tt hep-ph/0109045}}].
%%CITATION = HEP-PH/0109045;%%

\bibitem{Bauer:2002nz}
C.~W. Bauer, S.~Fleming, D.~Pirjol, I.~Z. Rothstein, and I.~W. Stewart, {\it
  {Hard scattering factorization from effective field theory}},  {\em
  Phys.~Rev.~D} {\bf 66} (2002) 014017,
  [\href{http://arXiv.org/abs/hep-ph/0202088}{{\tt hep-ph/0202088}}].
%%CITATION = HEP-PH/0202088;%%

\bibitem{Beneke:2002ph}
M.~Beneke, A.~Chapovsky, M.~Diehl, and T.~Feldmann, {\it {Soft collinear
  effective theory and heavy to light currents beyond leading power}},  {\em
  Nucl.~Phys.} {\bf B643} (2002) 431--476,
  [\href{http://arXiv.org/abs/hep-ph/0206152}{{\tt hep-ph/0206152}}].
%%CITATION = HEP-PH/0206152;%%

\bibitem{Catani:2013tia}
S.~Catani, L.~Cieri, D.~de~Florian, G.~Ferrera, and M.~Grazzini, {\it
  {Universality of transverse-momentum resummation and hard factors at the
  NNLO}},  {\em Nucl.~Phys.} {\bf B881} (2014) 414--443,
  [\href{http://arXiv.org/abs/1311.1654}{{\tt arXiv:1311.1654}}].
%%CITATION = ARXIV:1311.1654;%%

\bibitem{Catani:2011kr}
S.~Catani and M.~Grazzini, {\it {Higgs Boson Production at Hadron Colliders:
  Hard-Collinear Coefficients at the NNLO}},  {\em Eur.~Phys.~J.~C} {\bf 72}
  (2012) 2013, [\href{http://arXiv.org/abs/1106.4652}{{\tt arXiv:1106.4652}}].
%%CITATION = ARXIV:1106.4652;%%

\bibitem{Catani:2012qa}
S.~Catani, L.~Cieri, D.~de~Florian, G.~Ferrera, and M.~Grazzini, {\it {Vector
  boson production at hadron colliders: hard-collinear coefficients at the
  NNLO}},  {\em Eur.~Phys.~J.~C} {\bf 72} (2012) 2195,
  [\href{http://arXiv.org/abs/1209.0158}{{\tt arXiv:1209.0158}}].
%%CITATION = ARXIV:1209.0158;%%

\bibitem{Gehrmann:2012ze}
T.~Gehrmann, T.~L{\"u}bbert, and L.~L. Yang, {\it {Transverse parton
  distribution functions at next-to-next-to-leading order: the quark-to-quark
  case}},  {\em Phys.~Rev.~Lett.} {\bf 109} (2012) 242003,
  [\href{http://arXiv.org/abs/1209.0682}{{\tt arXiv:1209.0682}}].
%%CITATION = ARXIV:1209.0682;%%

\bibitem{Gehrmann:2014uaa}
T.~Gehrmann, T.~L{\"u}bbert, and L.~L. Yang, {\it {Transverse Parton
  Distribution Functions at Next-To-Next-To-Leading-Order}},
  \href{http://arXiv.org/abs/1401.1222}{{\tt arXiv:1401.1222}}.
%%CITATION = ARXIV:1401.1222;%%

\bibitem{Gaunt:2014cfa}
J.~Gaunt, M.~Stahlhofen, and F.~J. Tackmann, {\it {The Gluon Beam Function at
  Two Loops}},  \href{http://arXiv.org/abs/1405.1044}{{\tt arXiv:1405.1044}}.
%%CITATION = ARXIV:1405.1044;%%

\bibitem{Collins:1981uw}
J.~C. Collins and D.~E. Soper, {\it {Parton Distribution and Decay Functions}},
   {\em Nucl.~Phys.} {\bf B194} (1982) 445.
%%CITATION = NUPHA,B194,445;%%

\bibitem{Manohar:2006nz}
A.~V. Manohar and I.~W. Stewart, {\it {The Zero-Bin and Mode Factorization in
  Quantum Field Theory}},  {\em Phys.~Rev.~D} {\bf 76} (2007) 074002,
  [\href{http://arXiv.org/abs/hep-ph/0605001}{{\tt hep-ph/0605001}}].
%%CITATION = HEP-PH/0605001;%%

\bibitem{Bardeen:1978yd}
W.~A. Bardeen, A.~Buras, D.~Duke, and T.~Muta, {\it {Deep Inelastic Scattering
  Beyond the Leading Order in Asymptotically Free Gauge Theories}},  {\em
  Phys.~Rev.~D} {\bf 18} (1978) 3998.
%%CITATION = PHRVA,D18,3998;%%

\bibitem{Curci:1980uw}
G.~Curci, W.~Furmanski, and R.~Petronzio, {\it {Evolution of Parton Densities
  Beyond Leading Order: The Nonsinglet Case}},  {\em Nucl.~Phys.} {\bf B175}
  (1980) 27.
%%CITATION = NUPHA,B175,27;%%

\bibitem{Furmanski:1980cm}
W.~Furmanski and R.~Petronzio, {\it {Singlet Parton Densities Beyond Leading
  Order}},  {\em Phys.~Lett.~B} {\bf 97} (1980) 437.
%%CITATION = PHLTA,B97,437;%%

\bibitem{Ellis:1996nn}
R.~K. Ellis and W.~Vogelsang, {\it {The Evolution of parton distributions
  beyond leading order: The Singlet case}},
  \href{http://arXiv.org/abs/hep-ph/9602356}{{\tt hep-ph/9602356}}.
%%CITATION = HEP-PH/9602356;%%

\bibitem{Gatheral:1983cz}
J.~Gatheral, {\it {Exponentiation of Eikonal Cross-sections in Nonabelian Gauge
  Theories}},  {\em Phys.~Lett.~B} {\bf 133} (1983) 90.
%%CITATION = PHLTA,B133,90;%%

\bibitem{Frenkel:1984pz}
J.~Frenkel and J.~Taylor, {\it {Nonabelian Eikonal Exponentiation}},  {\em
  Nucl.~Phys.} {\bf B246} (1984) 231.
%%CITATION = NUPHA,B246,231;%%

\bibitem{Procura:2011aq}
M.~Procura and W.~J. Waalewijn, {\it {Fragmentation in Jets: Cone and Threshold
  Effects}},  {\em Phys.~Rev.~D} {\bf 85} (2012) 114041,
  [\href{http://arXiv.org/abs/1111.6605}{{\tt arXiv:1111.6605}}].
%%CITATION = ARXIV:1111.6605;%%

\bibitem{Cutkosky:1960sp}
R.~Cutkosky, {\it {Singularities and discontinuities of Feynman amplitudes}},
  {\em J.~Math.~Phys.} {\bf 1} (1960) 429--433.

\bibitem{Veltman:1994Ve}
M.~Veltman, {\em Diagrammatica : the path to Feynman rules}.
\newblock Cambridge University Press, 1994.

\bibitem{Becher:2008cf}
T.~Becher and M.~D. Schwartz, {\it {A precise determination of $\alpha_s$ from
  LEP thrust data using effective field theory}},  {\em JHEP} {\bf 0807} (2008)
  034, [\href{http://arXiv.org/abs/0803.0342}{{\tt arXiv:0803.0342}}].
%%CITATION = ARXIV:0803.0342;%%

\bibitem{Abbate:2010xh}
R.~Abbate, M.~Fickinger, A.~H. Hoang, V.~Mateu, and I.~W. Stewart, {\it {Thrust
  at N$^3$LL with Power Corrections and a Precision Global Fit for
  $\alpha_s(m_Z)$}},  {\em Phys.~Rev.~D} {\bf 83} (2011) 074021,
  [\href{http://arXiv.org/abs/1006.3080}{{\tt arXiv:1006.3080}}].
%%CITATION = ARXIV:1006.3080;%%

\bibitem{Binosi:2008ig}
D.~Binosi, J.~Collins, C.~Kaufhold, and L.~Theussl, {\it {JaxoDraw: A Graphical
  user interface for drawing Feynman diagrams. Version 2.0 release notes}},
  {\em Comput.~Phys.~Commun.} {\bf 180} (2009) 1709--1715,
  [\href{http://arXiv.org/abs/0811.4113}{{\tt arXiv:0811.4113}}].
%%CITATION = ARXIV:0811.4113;%%

\bibitem{Moch:2004pa}
S.~Moch, J.~Vermaseren, and A.~Vogt, {\it {The Three loop splitting functions
  in QCD: The Nonsinglet case}},  {\em Nucl.~Phys.} {\bf B688} (2004) 101--134,
  [\href{http://arXiv.org/abs/hep-ph/0403192}{{\tt hep-ph/0403192}}].
%%CITATION = HEP-PH/0403192;%%

\bibitem{Korchemsky:1987wg}
G.~Korchemsky and A.~Radyushkin, {\it {Renormalization of the Wilson Loops
  Beyond the Leading Order}},  {\em Nucl.~Phys.} {\bf B283} (1987) 342--364.
%%CITATION = NUPHA,B283,342;%%

\end{thebibliography}\endgroup

\end{document}